\documentclass{egpubl}
\usepackage{xcolor}
\usepackage{titlesec}
\usepackage{float}

\usepackage{fontawesome5}
\usepackage{subcaption}
\usepackage[normalem]{ulem}
\useunder{\uline}{\ul}{}
\usepackage{booktabs}
\usepackage{amsmath}
\usepackage{makecell}
\usepackage{lipsum} 
\usepackage{array}
\usepackage{longtable}
\usepackage{graphicx}
\usepackage{adjustbox}

\definecolor{clc}{HTML}{FFD541}
\definecolor{plc}{HTML}{9841FF}
\definecolor{rlc}{HTML}{FB7676}
\definecolor{alc}{HTML}{159DFF}
\definecolor{llc}{HTML}{9F9F9F}

\newenvironment{mydefinition}[1]{
    \noindent\colorbox{blue!20}{\textbf{#1}}\begin{itshape}\newline
}{
    \end{itshape}
    \medskip 
}

\newcommand{\LLmark}{\textcolor{llc}{\small \faSquare}~}
\newcommand{\CLmark}{\textcolor{clc}{\noindent \small \faSquare}~}
\newcommand{\PLmark}{\textcolor{plc}{\noindent \small \faSquare}~}
\newcommand{\RLmark}{\textcolor{rlc}{\noindent \small \faSquare}~}
\newcommand{\ALmark}{\textcolor{alc}{\noindent \small \faSquare}~}

\usepackage[T1]{fontenc}
\usepackage{dfadobe}  
\usepackage{algorithm}
\usepackage{algorithmic}

\usepackage{cite}  
\BibtexOrBiblatex
\electronicVersion
\PrintedOrElectronic

\usepackage{egweblnk}

\title{vi(E)va LLM! A Conceptual Stack for Evaluating and Interpreting Generative AI-based Visualizations}

\author[L. Podo \& M. Ishmal \& M. Angelini]
{\parbox{\textwidth}{\centering L. Podo$^{1}$\orcid{0000-0001-8780-6848},
          M. Ishmal$^{1}$ and M. Angelini$^{2,1}$\orcid{0000-0001-9051-6972} 
        }
        \\
{\parbox{\textwidth}{\centering $^1$Sapienza University of Rome, Italy\\
         $^2$Link Campus University, Rome, Italy
       }
}
}

\begin{document}


\maketitle
\begin{abstract}
The automatic generation of visualizations is an old task that, through the years, has shown more and more interest from the research and practitioner communities. Recently, large language models (LLM) have become an interesting option for supporting generative tasks related to visualization, demonstrating initial promising results. At the same time, several pitfalls, like the multiple ways of instructing an LLM to generate the desired result, the different perspectives leading the generation (code-based, image-based, grammar-based), and the presence of hallucinations even for the visualization generation task, make their usage less affordable than expected.
Following similar initiatives for benchmarking LLMs, this paper copes with the problem of modeling the evaluation of a generated visualization through an LLM. We propose a theoretical evaluation stack, EvaLLM, that decomposes the evaluation effort in its atomic components, characterizes their nature, and provides an overview of how to implement and interpret them. We also designed and implemented an evaluation platform that provides a benchmarking resource for the visualization generation task. The platform supports automatic and manual scoring conducted by multiple assessors to support a fine-grained and semantic evaluation based on the EvaLLM stack. Two case studies on GPT3.5-turbo with Code Interpreter and Llama2-70-b models show the benefits of EvaLLM and illustrate interesting results on the current state-of-the-art LLM-generated visualizations.
\end{abstract} 


 
\section{Introduction}

In the last year, Large Language Models (LLMs) have become everywhere across various disciplines, demonstrating remarkable efficacy and capabilities in performing different tasks. Noteworthy applications span from finance ~\cite{wu2023bloomberggpt} to coding ~\cite{chen2021evaluating}, exhibiting tangible impacts in disparate domains and presenting a valuable opportunity for human augmentation. For example, Github reports an enhancement in developer productivity by a 55\% increasing in writing code, attributed to the introduction of Copilot~\cite{GitHubCopilot}, an LLM model fine-tuned to generate code.
The relentless advancements in research contribute to the widespread usage of LLMs, thanks to the evolving proficiency in comprehending the underlying semantics in human instructions \cite{hadi2023large}. This results in diverse implications across different sectors, opening the research to new opportunities.
In the visualization field, LLMs exhibit promising capabilities in generating visualization as images and code \cite{wang2020generalizing, tian2023chartgpt}, using libraries such as D3.js~\cite{D3js} and Matplotlib~\cite{Hunter:2007}. A significant implication of this capability is that these models may empower non-expert users to generate insightful visualizations without prior data visualization expertise, offering a distinct advantage in creating visualizations through natural language queries~\cite{chen2022nl2interface, maddigan2023chat2vis}. In this direction, the LLMs find a potential fit in the field of Visualization Recommendation Systems~\cite{podo2023machine}, representing a valuable solution to the contemporary challenges posed by the large amount of data available in the different fields. These opportunities open a new research track: LLM 4 Data visualization (LLM4VIS). A promising use case is represented by OpenAI's CodeInterpreter, a sophisticated solution capable of generating visualization and delineating complex problems into logical steps for a more meaningful response~\cite{firdous2023openai}.
Despite the prevalence of models (e.g., GPT-4~\cite{openai2023gpt}, LLama~\cite{touvron2023llama}) and their continuous improvement, a significant portion of their behavior remains ripe for exploration and further scrutiny. 
To bridge this knowledge gap, researchers are investigating their capabilities across diverse benchmark datasets~\cite{wang2018glue, zellers2019hellaswag, clark2018think, zhong2023agieval}.
While natural language processing, general knowledge, common sense, problem-solving, advanced reasoning, and coding tasks have undergone thorough examination, visualization skills remain an area demanding further exploration due to its preliminary results. Although LLMs can generate visualizations, a comprehensive analysis of their visualization capabilities is lacking, prompting research questions such as \textit{Do LLMs adhere to visualization best practices?} or \textit{Can the LLM-generated visualizations be trusted?}.

In this paper, we cope with the problem of modeling the evaluation of LLM-generated visualizations and informing specific benchmarks for LLM-based visualizations to foster quantitative multi-faceted evaluation and comparability.
We introduce EvaLLM, a conceptual stack to evaluate LLM-generated visualization. It decomposes the evaluation effort in its atomic components, characterizes their nature, and provides an overview of how to implement and interpret their results. Moreover, we propose a user-friendly web-based platform that provides a benchmarking resource for the visualization generation task. The platform supports automatic and manual scoring and implements solutions for the EvaLLM composing layers and levels. 
In addition to automated assessments, the platform incorporates human-based evaluation features, supporting multiple users in identifying potential pitfalls or semantic errors. 
Finally, we present two qualitative use cases that evaluate GPT-3.5-turbo and Llama2-70b models on 50 samples from the NvBench dataset~\cite{luo2021nvbench} against the EvaLLM stack. 
The analysis of the use case results shows common errors in visualization generation, from the more structural to the more semantic errors, allowing their identification at specific levels of the EvaLLM stack. 
This evaluation not only showcases the capabilities of the proposed conceptual stack but also sheds light on the limits and pitfalls of GPT-3.5-turbo and Llama-70b in a practical setting.

Summarizing, the main contributions of this work are:
\begin{enumerate}
    \item EvaLLM: a conceptual stack for evaluating LLM capabilities in visualization generation task;
    \item a web-based platform for visualization evaluation providing an implementation of EvaLLM;
    \item two use cases involving GPT-3.5-turbo and Llama 70-b to show insights and opportunities obtainable by using EvaLLM.
\end{enumerate}

This paper is structured as follows: Section 2 reviews diverse prominent language model approaches and the research about their usage for visualization generation tasks. Section 3 introduces the proposed conceptual stack for assessing LLM-based visualizations and their composing layers. Section 4 illustrates possibilities for implementing EvaLLM levels of evaluation and their meanings. Section 5 discusses the design and implementation of a platform to instruct benchmarking activities leveraging EvaLLM. Section 6 shows the results of two use cases on GPT-3.5-turbo and Llama 70-b. Limitations and opportunities are presented in Section 7, with Section 8 concluding the paper.

\section{Related work}

A Visual Recommendation System (VRS) is an automated system that can generate insightful visualizations from a given dataset. As outlined by Podo et al.~\cite{podo2023machine}, VRS can be broadly categorized into task-agnostic and task-aware. The former relies solely on the dataset as input, requiring the system to learn how to generate insightful visualizations autonomously. In contrast, the latter involves user guidance through queries or utterances in addition to the dataset, facilitating a more interactive and user-informed approach to data visualization.
This section delves into the contributions coping with task-aware VRSs. It offers a comprehensive overview of how methods based on Large Language Models (LLMs) are employed and the potential enhancements they could bring. Furthermore, it explores also more classic approaches, including rule-based and machine-learning methods~\cite{setlur2016eviza, luo2021natural, srinivasan2023bolt, kim2021data, narechania2020nl4dv, song2022rgvisnet, gao2015datatone}. 
Finally, it addresses existing research investigating the capabilities and quality of an LLM-based VRS. 

\subsection{VRS task-aware benchmarks}
As discussed, the task-aware VRSs require user utterance and the dataset as input. 
To assess a model's performance on the given task, several datasets are available, each with its unique characteristics:
nvBench dataset~\cite{luo2021nvbench} presents 25,750 triplets (data, utterance, chart) sourced from 105 domains of tabular data. Despite its comprehensive nature, it is worth noting that many utterances in the nvBench dataset provide explicit visualization specifications, including exact details about the expected visualization, like the data columns, the chart type, or other visualization properties. For instance, the following example: "A pie chart showing the number of faculty members for each rank." precisely communicates the user's visualization expectations regarding the chart type. Moreover, it covers only the most frequent visualization types, such as bar, pie, line, scatter, stacked bar, grouping line, and grouping scatter.
The NLVCorpus~\cite{srinivasan2021collecting} comprises 893 utterances related to ten types of charts. However, the NLVCorpus is constrained by its reliance on only three data tables.
In the work by Tian et al.~\cite{tian2023chartgpt}, a new dataset is proposed based on the nvBench dataset. The authors curate this new dataset by initially selecting utterances involving only one data table, eliminating pairs requiring multiple tables. They then employ GPT-3 to abstract the utterances, thereby reducing explicitness. While this dataset is an essential resource in the field, it is important to note that it relies on synthetic pairs.
Quda~\cite{fu2020quda} encompasses 14,035 user utterance queries spanning various analytical tasks. However, it lacks associated charts, which limits its utility for visualization evaluation scenarios.
This diversity in datasets provides researchers with options for evaluating models, each dataset bringing its strengths and limitations. At the same time, none copes with the problem of defining a granular set of metrics and elements to consider for comparing the tested approaches. 

\subsection{Traditional visualization generation methods}
In the literature, the task-aware VRS models predominantly rely on traditional methodologies.
For example, in ~\cite{setlur2016eviza}, the authors introduce an interactive visualization tool that employs a hybrid approach integrating Natural Language Processing (NLP) and decision rules. This system combines a probabilistic grammar approach with predefined rules that can be dynamically updated based on the underlying data. This approach offers the advantage of reduced computational complexity compared to machine/deep learning methods, resulting in a faster interaction. 
In ~\cite{srinivasan2023bolt}, the authors present BOLT, a web-based platform for multi-dashboard authoring using natural language. The authors propose a system based on traditional NLP techniques to map user utterances to prevalent dashboard objectives and generate appropriate visualizations. 
The novelty of this paper lies in the focus on higher dashboard objectives that focus on the overall problem the user is trying to solve (e.g., change analysis) rather than lower ones that involve simpler visualization operations (e.g., sorting). Although user validation demonstrates the positive impact of natural language on interaction, the inherent challenge of handling ambiguous user utterances persists, posing difficulties for classical NLP methods.
In contrast, ~\cite{narechania2020nl4dv} introduces a method for interacting with a dataset based on NLP, focusing on generating a single visualization rather than suggesting a complete dashboard. The emphasis here is on providing tailored visualizations through NLP techniques.
Moving beyond rule-based approaches, ~\cite{luo2021natural} proposes ncNet, a seq2seq model that translates Natural Language Queries (NLQs) into a custom visualization grammar, Vega-Zero. This transformer-based encoder-decoder architecture employs an attention-forcing method and is trained on the nvBench dataset. While ncNet represents a breakthrough in processing the user inputs as free text, it still faces challenges in handling ambiguous and ill-posed natural queries.
To address these limitations, ~\cite{song2022rgvisnet} introduced RGVisNet, a retrieval and generation-based approach. Instead of directly translating NLQs into the visualization grammar, the network is divided into two components: (i) a data visualizations query retrieval network and (ii) a data visualizations query revision network. Based on a Graph Neural Network (GNN), the retrieval network retrieves relevant data visualizations from a database. In contrast, the revision network adjusts the candidate based on the expected data visualization outcomes through a decoder structure.
Despite the effectiveness of these non-LLM-based systems, they grapple with the challenge of capturing underlying semantics in user utterances, mainly when dealing with ambiguous expressions. This may not lead to an ideal generated visualization as a result.

\subsection{LLM-based visualization generation methods}
As discussed, the rise of the LLMs has posed new research questions in different fields, including the data visualization field. New approaches have been proposed in this direction. 
In \cite{hong2023conversational}, the authors propose AI Thread, a chatbot for multi-threaded analytic conversations. Using the chain-of-thoughts reasoning technique \cite{wei2022chain}, the system leverages GPT-3.5 capabilities in Python to map the user utterance into a visualization using Matplot and Seaborn libraries. Moreover, the tool proposes a two-view interface: the main chat, that is, the chat view interaction, and the threaded panel that allows for edits on the visualizations in the main board. A different approach is proposed in \cite{chen2022nl2interface}. The authors discuss a method based on few-shot learning~\cite{wang2020generalizing} at inference time on the Codex LLM model by OpenAI~\cite{finnie2022robots}. The model is fed with natural language-SQL (NL2SQL) pairs examples and the user's natural language query to aid in task understanding. The result is then converted into Vega-Lite specifications using a rule-based approach~\cite{chen2022pi2}. The authors emphasize that their approach can achieve valuable results without requiring model tuning. However, the study's evaluation is conducted on a limited number of examples without providing any findings about the quality of the explored samples, raising questions about the generalizability of the approach.
Similarly, Maddigan et al.~\cite{maddigan2023chat2vis} present a comparable study involving Codex, GPT -3, and ChatGPT. Like the previous work, the study lacks a comprehensive discussion of results and relies on a small number of evaluation samples. 
Finally, Tian et al.~\cite{tian2023chartgpt} recently introduced ChartGPT, a multi-step pipeline incorporating LLMs into various stages, breaking down the visualization generation problem into logical steps. The authors fine-tune FLAN-t5~\cite{chung2022scaling} to align the model with the intended task. The evaluation is extensive in the number of tests but is still executed using a custom evaluation scheme. This problem is common even to other papers presented in this section, leading us to study deeply how a general framework could support LLM-generated visualizations.

\subsection{Evaluating LLMs visualizations}
Despite the impressive performances demonstrated by these new approaches, there needs to be more evaluation approaches and standard methodologies to assess the understating capabilities in the data visualization domain. To fill this gap, some researchers have started to map these capabilities, proposing evaluation methods. 
For example, \cite{chen2023beyond} presents an evaluation study focusing on GPT-3.4 and GPT-4, examining these models beyond the conventional visualization generation task considered from just the code generation perspective. Instead, the study delves into multiple facets, such as data interpretation, visualization design, visual data exploration support, and insight communication. To assess the model's proficiency, they tasked GPT with completing various quizzes and producing visualizations based on the data visualization course of Harvard CS171. GPT was asked to perform random tasks from the dataset, including quizzes and homework, and a group of fellows evaluated the results. This study introduces an initial evaluation approach to understand how well GPT can address challenges closely tied to visualization tasks. While using general categories of elements considered during a visualization evaluation, the work focuses more on their application than the conceptual modeling what it means to evaluate an LLM-generated visualization. It also presents interesting considerations on this topic, and they inspired us to define our evaluation conceptual stack.
Another in-depth study is conducted by Kim et al.~\cite{kim2023good}, where the authors thoroughly explore the capabilities and limitations of ChatGPT in visualization tasks. They pose a series of questions from the VisGuides forum and compare the answers with human responses from the original questions on the blog. The study observes that ChatGPT performs similarly to human responses and, in some cases, even outperforms them. The authors then asked a group of data visualization experts to evaluate visualizations provided by both ChatGPT and domain experts without revealing the source. The study concludes that users prefer human feedback over LLM feedback, highlighting the main disadvantages of ChatGPT's lack of discussion and emphasizing the potential advantage of a design knowledge agent. We identified similar needs in including human assessment even for LLM-based visualizations, which informed the creation of EvaLLM.
Finally, Tao et al.~\cite{tao2023mapping} investigate ChatGPT's capability to generate abstract maps through two main experiments. The first experiment assesses the model's ability to create thematic maps based on specified map styles and data sources in the prompts. The second experiment involves generating mental maps from the data. The study concludes that ChatGPT can assist in developing maps and enhance productivity. Nonetheless, it underscores the model's dependence on external tools for visualization generation, highlighting challenges in achieving the desired output.
While these works contribute to integrating LLMs into existing data visualization pipelines, few have systematically studied the main characteristics needed to evaluate a generated visualization and how to organize and leverage them to benchmark automatic generators, creating a taxonomy of recurring errors. 
To fill this gap, this paper proposes a conceptual stack that includes automatic quantitative and human-based evaluation metrics and investigates its application to LLM-based visualizations.


\section{EvaLLM: Characterizing the evaluation of LLM-generated visualizations}
\label{sec:stack}

\subsection{Preliminaries}
\label{sec:preliminaries}

In the domain of LLMs applied to visualization generation task (LLM4VIS), the challenge is represented by creating a solution capable of accurately generating a visualization given a dataset $\mathcal{D}$ and a user query $\mathcal{Q}$.
In the forthcoming discussion, it is crucial first to provide a formulation of the problem, as formalized in the following definition:

\begin{mydefinition}{Problem definition}
    Given a tabular dataset $\mathcal{D}$ of $n$ features $\Phi = \{\phi_1,\dots,\phi_n\}$ and a user query $\mathcal{Q}$, an LLM-based visualization recommender system, generates a visualization $V$ that provide the best accuracy concerning $\mathcal{Q}$.
\end{mydefinition}

A task-aware LLM-VRS should be able to discern and understand the user's intent as conveyed in the input query and generate a visualization that closely aligns with the user's expectations.
Building upon the existing literature for interactive visualizations in this domain, we can categorize the output visualization $\mathcal{V}$ into three classes: image-based visualization, code-based visualization, and grammar-based visualization. In the first case, the visualization is directly generated as an image, representing the only output of the generation process. A code snippet is generated and then used to render the visualization in the code-based case. Rendering such output necessitates a specific environment for code execution relative to the language of the generated code. Conversely, the grammar-based case relies on visualization grammars, such as VegaLite, enabling direct rendering by mapping the grammar with one or multiple code environments.
Based on these assumptions, we propose EvaLLM, a conceptual stack that allows the evaluation of all these classes of generated visualizations.

\subsection{Structuring the evaluation: the EvaLLM layers}
\label{se:layers}

\begin{figure*}[t!]
    \centering
    \includegraphics[width=0.9\textwidth]{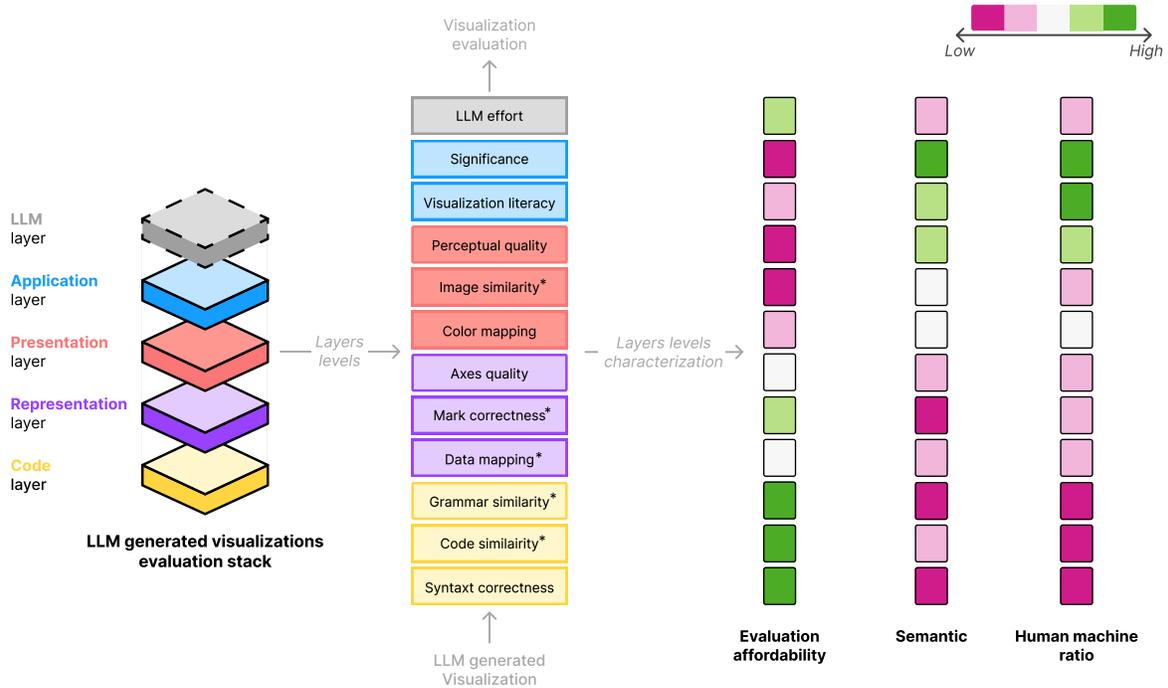}
    \caption{EvaLLMStack: concept evaluation stack for evaluating the LLM-generated visualizations}
    \label{fig:stack}
\end{figure*}

EvaLLM is a conceptual stack to model the evaluation for LLM-generated visualization, as shown in Figure~\ref{fig:stack}. Due to the early stage of the LLM models for the visualization field, there is a lack of standard methods for assessing the quality of generated visualization, from what their quality depends, or which design elements of visualization the LLMs are better at accurately generating. Drawing inspiration from the ISO/OSI model~\cite{day1983osi}, EvaLLM involves abstract layers to evaluate specific visualization properties and derive a corresponding quality measure.
From bottom to top, each layer transitions towards a higher level of abstraction. 
As illustrated in Figure~\ref{fig:stack}, EvaLLM comprises five primary layers. 
Each layer is subdivided into levels, where the focus is directed explicitly toward assessing distinct visualization properties of the same layer. The rationale for this structure is to allow for an initial set of homogeneous categories specialized internally in further detail inside each of them (through levels) and support in this way a fine-grained evaluation of the generated visualization properties, better characterizing the LLMs' expressive power.

At the base of the EvaLLM stack lies the \textbf{Code layer}, which plays a role in evaluating the fundamental structural properties of a visualization.
The primary question that the Code layer seeks to address is \textit{RQ.1: "Is the visualization consistent within the code environment it is created with?"}. This inquiry delves into the coherence of the visualization within the structure of the related environment (e.g., a generic programming language, a specialized one, or a grammar), checking whether the generated code aligns with the expected structure or evaluating which differences may be present. These differences are mostly syntactical, but they can already present interesting cases that strongly affect the final result quality (i.e., different syntactic structures that give birth to similar visual results or slightly different codes that produce very different results). In other words, the Code layer focuses on verifying the syntactical correctness and integrity of the visualization's underlying code, laying the groundwork for subsequent layers to delve into more nuanced aspects of evaluation.
Immediately above the Code layer, the visualization evaluation is moved to the \textbf{Representation layer}. At this stage, the focus shifts to semantic aspects of the visualization representation, focusing on the core properties of the visualization related to the representation rules defined by the literature~\cite{spence2001information, munzner2009nested}. We distilled from the literature three prominent aspects: the data mapping rules, the choice of the appropriate visual encodings (i.e., marks), and the representation properties of eventual axes or reference visual elements. 
The evaluation question behind this layer is \textit{RQ.2 Are the data representation rules correctly generated in the visualization?}.
We notice that already at this layer, it is not granted that some of these aspects are directly captured in the user query (i.e., the query may or may not include a direct reference to these aspects). At the same time, they are an integral part of a correct and accurate visualization. For this reason, at this layer, a more fine-grained and quantitative evaluation is needed concerning a coarse one based only, for example, on evaluating the correctness of the visual encoding. This evaluation should focus on quantitative distances between the generated and expected content.
On top of this layer, there is the \textbf{Presentation layer}. This layer is the third layer in the stack, and it assesses the data presentation quality of the visualization from the perceptual standpoint~\cite{ware2019information}. This layer aims to model the design choices for perceptual aspects made by the LLM on the generated visualization to answer the question \textit{RQ.3 Is the visualization correctly presented, comprehensible by a human user, and not giving a perceptual error or illusions hindering the human interpretation of its content?}. For example, it evaluates aspects such as the quality and appropriateness of the color mappings, the distinguishability of visual elements, or the perceptual comprehensibility of the visualized information.
The fourth layer of the stack is the \textbf{Informativeness layer}. It is responsible for measuring the more intrinsic quality of the visualization in terms of its insightfulness and adherence to best practices in visualization literacy, answering the question \textit{RQ.4 Is the visualization insightful, and how well it supports the user in answering its information needs?}. This involves an evaluation that examines the capability of the visualization to convey meaningful insights concerning the user query and to be aligned with best practices of visualization literacy to ease the visualization understanding by the user as much as possible.
Lastly, the last layer on top of the stack is the \textbf{LLM layer}. 
The LLM layer contributes twofold at the end of the stack. Its first goal is to evaluate the strategy for generating the specific visualization. Possible choices relate to single prompting the LLM, applying prompt engineering~\cite{white2023prompt}, or more sophisticated strategies like Chain-of-thought~\cite{wei2022chain}. Evaluate the configuration elements to discern between different models (i.e., plain foundation model, zero-shot/few-shot approaches, fine-tuning, or a trained-from-scratch new model). In this way, it is possible to evaluate the effort for the generation process on top of the quality of the generated visualization. The second goal is evaluating the visualization by assessing its significance and adherence to best practices in the visualization literature, but this time, trying to measure the insightfulness of a visualization regarding the LLM knowledge and answering the questions \textit{RQ.5 What is the cost of the generated visualization?} and \textit{RQ.6 Is the visualization insightful, based on the LLM knowledge?}. 

\section{Implementing layers: the EvaLLM levels}
\label{sec:levels}

While the EvaLLM layers provide the overall structure of the fine-grained evaluation supported by the stack, they are still too coarse to be implemented. For this reason, EvaLLM presents a set of levels for each layer that better supports the implementation of the evaluation process and specifies how to interpret the evaluation results for each of them.

In this section, we delve into the exploration of such levels using the strategy of \textit{What, Why and How} - \textit{what} is the scope of the level, \textit{why} it is essential and \textit{how} it could be implemented.
Each level is characterized by three dimensions, orthogonal to all the levels, and necessary to describe the interpretation of evaluation results: \textbf{evaluation affordability, semantics, and human-machine ratio}.
\textit{Evaluation affordability} serves as a qualitative measure of the concrete feasibility of each level and the trust that can be put in the obtained indicator. For instance, the syntax correctness level is high on the proposed scale as shown in Figure \ref{fig:1}, because it is straightforward to implement and presents a clear-cut meaning, leaving little space for interpretation. On the other hand, the significance level could be implemented in multiple ways and leave a large space for interpreting its results; for this reason, it offers a low score on this aspect. \textit{Semantic} defines the degree to which an EvaLLM level provides a semantic evaluation of the visualization. For example, the syntax correctness level provides a minimal semantic evaluation, limiting itself to express if the generated code is syntactically correct or not, compared to the visualization literacy level, which instead provides a much higher level of semantics for evaluating how much a generated visualization is compliant with visualization best practices (i.e., the visualization could be accurately generated, but its efficacy for the user could be hindered by not following visualization best practices).
Finally, \textit{Human-machine ratio} gauges whether a level undergoes a fully automated evaluation process, relies on human-based assessment, or a mix of the two.
The 5-level Likert rating scale ranges from low (violet) to high (dark green), with intermediate levels of medium-low (pink), medium(white), and medium-high (light green), offering a qualitative overview for each EvaLLM level.
For the sake of clarity, we categorize the levels into two main classes: the one denoted by the asterisk (*) in the name requires a ground truth or a reference visualization to provide the quality measure against the generated one; differently, layers lacking this symbol entail quality metrics that do not rely on a predefined reference.

\subsection{\CLmark Syntax correctness level}
\textit{What.} The syntax correctness level is designed to verify whether or not an LLM-generated visualization is consistent within the syntax structure it is created with, i.e., code-based or grammar-based, and it is executable in the related environment to render the visualization. \textit{Why.} As it can be deducted, this level is pivotal in the stack to start evaluating a visualization. If the visualization code or the generated image has some structural errors and cannot be rendered, all the other levels are disabled in the evaluation process, and it stops here. \textit{How.} Considering a grammar-based visualization, e.g., VegaLite, the syntax correctness level verifies that the generated visualization specification respects the grammar rules and, subsequently, can be executed. An additional check concerns the capability to render a visualization (e.g., the grammar specification could be correct for the data part but not presenting a rendering part).
For instance, if the grammar choice is VegaLite, a straightforward implementation could involve the Altair library \cite{altairAltairDiscover} to verify the correctness of generated specifications. Differently, if the grammar is VegaZero \cite{luo2021natural}, the generated grammar could be verified using the method of sanity check proposed by the authors.
For this level, the affordability is high (green) because the expected result can be shaped as a binary result - correct or not, leaving no space for the interpretation of the evaluation. Differently, the semantic dimension is low (violet) since this level involves only syntactic visualization properties. Like the previous dimension, the human-machine ratio is also low (violet), meaning these checks can be executed automatically.

\subsection{\CLmark Code similarity level}
\textit{What.} The Code Similarity level processes the visualizations from the previous level, treating both the generated and the ground truth as code snippets and evaluating their similarity.
\textit{Why.} This step is needed as, most of the time, the same or similar visualizations can be generated through quite different code constructs. On the other hand, small code changes may produce big changes in the final visualization. LLMs are tested at this level for their capability to construct a visualization code similar to what a human user would do or to evaluate the differences and eventually the reasons behind them (e.g., better generalizability, better usage of coding practices, more efficient code).
\textit{How.} Haq and Caballero~\cite{haq2021survey} propose an extensive review of more than 70 binary code similarity approaches that can be leveraged for this level depending on the chosen code environment. Another example in the coding realm that can be transposed in the visualization one is discussed in \cite{liu2018supporting}. The authors propose a method to compare two code snippets to get commonalities and differences between them. Then, this information is visualized using a graph highlighting the relationships between the two codes. Finally, in such a context where slight variations in syntax can lead to semantically equivalent outputs, fuzzy matches like BLUE \cite{doddington2002automatic} could overcome the problem of the other rigidity in comparing the code.
In this case, the human-machine ratio is rated low, as the process can be entirely automated. On the semantic ratio, the score is medium-low as, while on the pure similarity, there is not much semantic involved, some semantic aspects can be captured by the patterns in which the LLM generates a specific visualization code. Analyzing them could improve the overall capability of the LLM to align with human specifications. Finally, the affordability is scored as high, as the structural construction can provide an initial level of assessment of the characteristics of the visualization (e.g., identifying missing code for specific parts of the generated visualization) and support a qualitative analysis of the compliance of the generated code with the ground truth.

\subsection{\CLmark Grammar similarity level}
\textit{What.} This level is tasked with measuring the similarity of the generated visualization's grammar structure compared to the ground truth structure. This level focuses mainly on the represented structure and values and less on the single identifiers given to the structure keys. \textit{Why.} The primary goal is to assess how effectively the model translates user query requests into a correct grammar structure. This structure should firlty align with the expected structure within the chosen grammar. Then, the assessment involves a comparison with the ground truth to check for efficiency in representation. Additionally, it aims to highlight structural differences in how a Language Model (LLM) represents the user query in the given grammar, comparing the results with the human-provided ground truth or with other LLMs.. \textit{How.} A conceivable implementation utilizing a JSON-based grammar would focus on comparing only the keywords of the generated grammar with their ground truth counterparts, omitting consideration of the values assigned to the keywords, which are reserved for higher levels. A practical implementation is represented by Playwright~\cite{playwright} that easily allows the comparison of JSON structures and highlights possible differences. The affordability of this level is rated as high because the process results in a numerical similarity score between syntactical structures. At the same time, human analysis can isolate blocks for which, for example, the same semantic is captured but with very different structural forms. Furthermore, this process involves semantic analysis only for structural differences, leading to a medium-low semantic score. Finally, the human-machine ratio is set to medium-low, given the fully automated process that could be employed to generate the similarity score, with a human intervention only in analyzing the differences.

\subsection{\PLmark Data mapping level}
\textit{What.} This level measures how well the generated visualization encodes the right data from the dataset compared to the ground truth. It assesses whether the columns chosen from the dataset align with the ground truth and if the model adeptly maps them according to their correct types, for example, ordinal or temporal. \textit{Why.} The main challenge is represented by ambiguous queries that could lead the model to select incorrect columns from the dataset, encode them in the wrong types on the axes, or hallucinate and select non-existing columns. \textit{How.} A possible metric is the data axes accuracy proposed by Podo et al.\cite{podo2023machine}. The authors propose to use the accuracy defined as the number of data columns correctly predicted on the axes over all the instances evaluated. A similar approach to the previous, taken from the text-to-SQL field, is the Query match accuracy \cite{xu2017sqlnet}. This metric considers a prediction correct only if all the expected matches, axes values and types, are correct. The main limitation of this metric is related to the zero score problem. To overcome this problem, in \cite{hazoom2021text}, the authors propose the Partial Component Match F1. 
Unlike the previous accuracy metric, this metric considers each data mapping match alone and calculates the F1 score. The final PCMF1 score of a predicted output is actually the average F1 score of all the single matches. 
An other approach, specifically designed for D3, is proposed in \cite{hull2023visgrader}. The authors propose an automatic approach for evaluating data
bindings, visual encodings and other properties of a D3 visualization that could be involved to return a quality measure of a D3 LLM generated visualization.
For this level, the affordability is medium due to the multiple data columns to consider and the possibility of defining a distance between different data types and evaluating the errors in non-binary form. Unlike affordability, semantics is scored medium-low since the data mapped conveys data semantics and analysis goals in isolation (i.e., single data columns) and in combination (e.g., correlation, trend analysis, etc.).  
Finally, the human-machine ratio is also medium-low. While the process is mainly automatic, it could benefit human evaluation to verify the resulting data mapped into the visualization, as discussed in \cite{katsogiannis2023survey}.

\subsection{\PLmark Mark correctness level}
\textit{What.} This level assesses the similarity between the visualization mark of the generated output and the corresponding ground truth and its usage of the chosen mark(s). 
\textit{Why.} When a model produces a visualization, errors extend beyond simple mislabeling of the mark type, such as mistaking a bar for a line. The model may also introduce errors in how the mark is used in the overall structure of the visualization, creating what we labeled ``visual hallucinations''. An example is illustrated in Figure~\ref{fig:gpt_grid}-g, where the model correctly recognizes the bar mark as the one to use. Still, it fabricates an uncommon and not usual representation of a bar chart.
\textit{How.} A potential implementation could adopt a hybrid methodology: automated checks could verify the accuracy of the mark while identifying hallucinations might necessitate human scrutiny. Consequently, affordability is rated as medium-high, primarily because the evaluation of the used mark helps evaluate the correctness of the visual representation. At the same time, its wrong usage could hinder the quality of the final outcome. Similarly, the human-machine ratio is medium, as human reviewers may be needed and should manually annotate eventual strange usage of a well-known mark, describing its characteristics. On the other hand, the semantic aspect is medium-low, given that the analysis relies on syntactic properties.

\subsection{\PLmark Axes quality level}
\textit{What.} The Axes Quality Level is structured to assess the quality of the axes' properties in a visualization.
\textit{Why.} The efficacy of a visualization hinges on its ability to clearly convey information to the users, with the axes playing a pivotal role. Optimal selection of these properties, such as axes orientation, scale, and ticks selection, is essential for delivering meaningful insights to users.
\textit{How.} One plausible implementation strategy entails an evaluation approach that compares the axes of the generated visualization with its ground truth. Alternatively, a set of rules defined by domain experts could be employed to assess the quality of the axes. For more complex axes properties, like axes arrangement, different methods and complexity could potentially be used for axes-based representation like parallel coordinates, radar chart, Radviz, or star coordinates.
For this level, we assign a medium affordability score since the diverse approaches used can influence the interpretation of the results. Conversely, both the human-machine ratio and semantics are rated low. The assessment of axes quality relies predominantly on syntactic properties, and the evaluation process is mostly automatic.

\subsection{\RLmark Color mapping level}
\textit{What.} This level examines the efficacy of color usage in encoding data attributes to convey data features.
\textit{Why.} The selection of a color palette depends on the nature of the data being represented. Evaluating how the model employs colors based on the data type and the chosen mark is pivotal for effective visualization and correct user interpretation.
\textit{How.} For example, Szafir et al.~\cite{szafir2017modeling} provide a set of perceptual data results from crowd-sourced studies that could be used to create probabilistic models to provide support and evaluate the color properties of the visualizations. Another possible implementation is discussed by Liu et al.~\cite{liu2018somewhere} that, based on a set of experiments, provide methods to generate color recommendations that could be used to design quantitative color evaluation metrics. In Szafir et al.~\cite{szafir2016visualization}, the authors propose a framework for optimizing the color choices in a data visualization that could be used to implement quantitative measures.
The type of implementation significantly influences the process, leading to a medium rating for both the human-machine ratio and semantics. Users are tasked with characterizing results in this domain. In contrast, affordability is marked as medium-low, as the interpretation of results heavily connotates visualization quality.

\subsection{\RLmark Image similarity level}
\textit{What.} This level involves a pixel-level comparison between the generated visualization and the ground truth.
\textit{Why.} Conducting a pixel-level comparison abstracts the assessment of the generated visualization from the specific generating environment or evaluates its characteristics by the final generated image, offering a direct and perceptual evaluation of how well the image aligns with the ground truth.
\textit{How.} An effective strategy entails applying computer vision techniques to quantify the structural similarity between the images, such as SSIM \cite{rouse2008understanding} or LPIPS \cite{kettunen2019lpips}. A different approach involves using the Siamese neural network \cite{appalaraju2017image} for image comparison.
Similar to the previous level, the affordability of the results is low, warranting a medium rating for similar reasons. The medium-high score also holds for the semantic and human-machine ratio, emphasizing the importance of the strategy in play.

\subsection{\RLmark Perceptual quality level}
\textit{What.} This level examines the visualization from a perceptual standpoint, focusing on ensuring that all perceptual properties are effectively encoded in the visualization and that it does not break any of the perceptual rules that are listed in visualization research (e.g.,~\cite{ware2019information}). \textit{Why.} Leveraging principles of visual perception, such as position along a common scale, length, direction, angle, area, and color hue, can help the visual interpretation from a human user and design more informative graphics. It is important to consider the effectiveness of visual encoding, ensuring that the importance of the attribute matches the salience of the channel used for them and avoid perceptual pitfalls like watchdog effects or usage of wrong visual channels.
\textit{How.} The control could be based on a stop list of perceptual pitfalls, eventually organized by visual marks, to be tested against the generated visualization to check for their presence automatically. On the contrary, assessing a human assessor is crucial at this level, as it can help identify more high-level problems faster.
We have characterized this level as low for affordability as it can describe aspects of the visualization that strongly affect its final quality. At the same time, some of the proposed methods could introduce uncertainty in their evaluation and require human intervention. The same score is assigned to the Human-machine ratio (predominance of human judgment over automatic approaches) and semantics.

\subsection{\ALmark Visualization literacy level}
\textit{What.} This evaluation level seeks to determine whether the model adhered to best practices in visualization literacy while generating the visualization. \textit{Why.} After the model produces a visualization that excels in other previous evaluation criteria, it does not automatically imply optimal configuration. The visualization could be requested to comply with best practices tailored to fit the task or particular analysis on top of the visual representation and presentation choices.
Therefore, assessing how the model incorporates best practices in generating the visualization is crucial. \textit{How.} In this direction, different works have been proposed in the literature, such as the work by Boy et al.~\cite{boy2014principled} that proposes a series of visualization best practices for line charts, bar charts, and scatterplots. Another possible approach is discussed by Lee et al.~\cite{lee2016vlat} that introduces a visualization literacy assessment test (Vlat) exploitable for assessing the compliance of LLM-based visualizations. Finally, an other implementation is proposed in \cite{joshievaluating}. The authors propose an evaluation study of LLMs on Parallel Coordinate Plots, evaluating how well the models were able to adhere to the based practices, based on Bloom’s taxonomy.
We classify the Semantic dimension as medium-high, as it demands a significant level of semantic understanding. While certain semantic aspects can be captured through patterns based on expert design rules, its overall complexity leans towards the higher end. Affordability is rated medium-low, primarily attributed to the absence of a well-defined output structure. Lastly, the human-machine ratio is set to high, necessitating the involvement of a human evaluator for effective processing.

\subsection{\ALmark Significance level}
\textit{What.} The significance layer represents the layer in the EvaLLM to measure the insightfulness~\cite{Battle2023} of a visualization. \textit{Why.} The significance is essential for analyzing complex data, identifying patterns, and extracting valuable insights. By simplifying complex information and presenting it visually, decision-makers can make informed and effective decisions quickly and accurately.
\textit{How.} As the visualization literacy level, this level should be implemented by involving a human reviewer using a dedicated platform. Performing this task automatically is extremely complex because the insightfulness lacks a mathematical formulation that could make the evaluation automatic. Interestingly, very recently, some work emerged in the Visual Analytics literature trying to mathematically formulate the relation between task support and insights generation, such as Suh et al.~\cite{suh2023grammar} for hypotheses testing using a visualization system or the works by North et al.~\cite{North2006,North2011} looking at most established literature.
We have characterized this level as high for semantics and human machine ration dimensions. The considerable degree of interpretation required for the visualization assessment necessitates human intervention, and the outcome is inherently semantic. Unlike the previous, the output lacks a clearly defined structure, as seen in binary or quantitative evaluations, contributing to low affordability on the scale.

\subsection{\LLmark LLM effort}
\textit{What.} This level focuses on assessing the effort of generating a visualization considering computational and methodological factors. Looking at the former, such as the computational costs, the inference time (real-time versus quasi-real-time), and the size of the models, they return a quantitative assessment of the efficiency of the generative model.
Looking at the latter, we list plain prompting (e.g., a single prompt representing the full user query, as in the case of the queries present in the NVBench), degree of prompt engineering, Chain-of-thoughts \cite{wei2022chain} (where the user split in a strategic way the sequence of prompts, enriching it with examples and counter-examples, to get a better-fit result)
or by chaining results of multiple models or the same model multiple times. These different strategies present an increasing effort to be implemented, having as benefits potentially more precise results. The government of this trade-off between raised precision and incurring cost is what this level aims at measuring.
\textit{Why.} Generating a visualization using an LLM is not only a matter of visualization quality but also of the strategy's performance and costs. For instance, even if using the same model with two different learning strategies could still generate the same expected visualization, the computational costs could be extremely different. Involving a zero-shot strategy is much computationally cheaper than fine-tuning a model, resulting in a minor effort.  
Hence, it cannot be concluded that one approach is better than another based only on the visual quality of the output, because also the effort is crucial to be considered.
\textit{How?} This level could be developed as a scoring function considering all the factors described to return an effort score. Based on the computational effort required, it could provide a normalized score - 0 low effort, 1 high effort.
For example, if we consider the learning strategies, zero-shot would have a lower effort and consequently a low value, compared to fine-tuning, which requires a higher effort, resulting in a higher score. 
However, evaluating the effort associated with quantifying methodological factors poses challenges. For instance, generating an image with CoT may entail a different effort than using chaining methods. Moreover, even if the methodology remains constant, its implementation can vary among users, making automatic quantification more challenging.
In light of these complexities, we assign a medium-low score for semantics. It's important to note that this score isn't a standalone numerical representation of effort but must be interpreted in the context of non-standard methodologies that differ based on user implementation. This nuanced approach also applies to the human-machine ratio, where human intervention may be necessary to quantify the effort involved in the user's developed methodology.
Ultimately, the affordability score is deemed medium-high due to the potential for human intervention in estimating the method, as opposed to automatically quantifying the effort for computational costs.

\section{LLM as evaluator}
\label{sec:agents}

\begin{figure}[t!]
    \centering
    \includegraphics[width=0.85\linewidth]{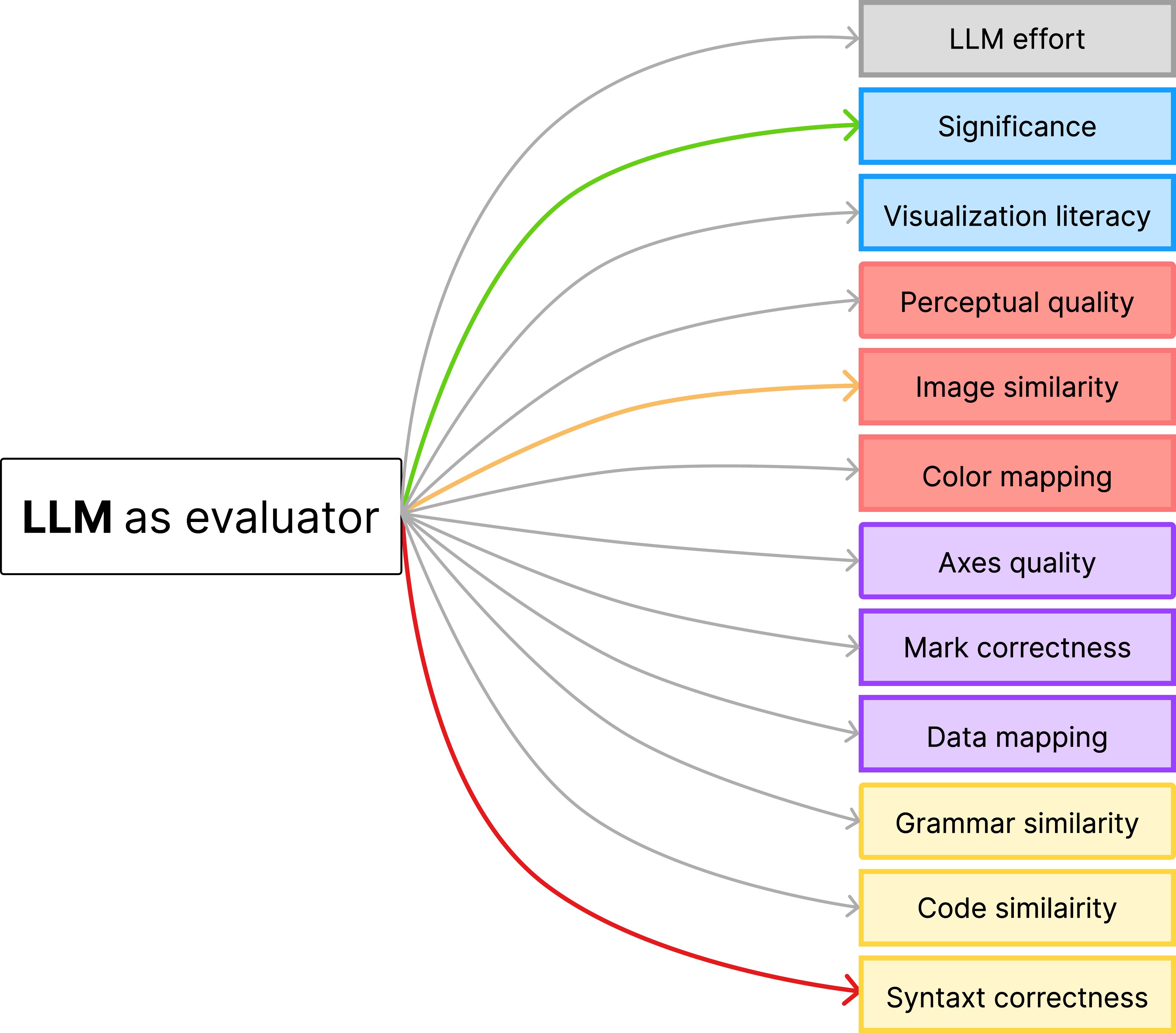}
    \caption{TThe image depicts the LLM's potential involvement at each level, with colors indicating impact in the three scenarios discussed. Red signifies suboptimal integration, yellow prompts further evaluation, and green denotes positive impacts.}
    \label{fig:llm agents}
\end{figure}

In the previous section, we have proposed the implementation of evaluation operations through a reasoned mix of automatic approaches and human involvement. For instance, we advocate for a fully automatic evaluation using a deterministic approach at the Syntax Correctness level. Conversely, for the Significance level, we have highlighted how human evaluation is better for assessing the visual quality of the presented data than automatic approaches.
While existing literature presents various automatic approaches, they are limited to a fixed number of EvaLLM levels and present more difficulty in being applied for the higher levels in the stack. Conversely, human evaluators can cover them, but their activities introduce not negligible cost in terms of time and effort needed.
In this context, we identify a new opportunity by involving the LLM as an additional evaluator, presenting a favorable trade-off between automatic and human evaluation. The LLM is orthogonal to all EvaLLM levels and represents an opportunity for implementing or supplementing existing approaches and for evaluation techniques, as discussed in recent contributions in the literature~\cite{chiang2023can, sun2023principle, fu2023gptscore}. 
However, while the LLM can potentially contribute to each level, a nuanced evaluation of its application at different levels becomes crucial, identifying the best fit, risks, and opportunities (see Figure~\ref{fig:llm agents}). For instance, introducing the LLM at the Syntax Correctness level could result in an over-engineered solution, adding complexity and increasing evaluation costs for a task more efficiently addressed by classic automatic techniques. Differently, involving an LLM at the Significance level, where automatic approaches face challenges and human evaluation is costly and time-consuming, represents a valuable opportunity, to complement with risk indicators and explanations of the LLM decisions. 
In the middle of these two extremes, we identify best-fit scenarios: for example, considering the image similarity level, assessing the benefits against costs is essential: despite the LLM's capacity to interpret images, it is critical to quantify its advantages against alternatives such as diffusion models or traditional computer vision techniques.

Exploring this new prospect lies beyond the scope of this paper, not affecting the EvaLLM stack definition but complementing its implementation with a new possibility, that will be the subject of future research activities. 

\section{EvaLLM assisting evaluation platform}
\label{sec:platform}

\begin{table*}
    \centering
    \renewcommand{\arraystretch}{1.5} 
     \begin{tabular}{|>{\arraybackslash}m{2.5cm}|>{\arraybackslash}m{5cm}|>{\arraybackslash}m{7.5cm}|>{\arraybackslash}m{1cm}|}
            \hline
            \textbf{Level} & \textbf{Description} & \textbf{Formula(s)} & \textbf{Human-Ratio} \\

            \hline
            \textit{Syntax  \newline Correctness} &
            Outputs 1 if the visualization is rendered successfully, 0 if not. &
            - &
            \faCogs \\

            \hline

            \textit{Code \newline Similarity} &
            Returns the similarity between G.T and Generated Viz. in terms of code plagiarism &
            PyCode Similarity Library &
            \faCogs \\ 

            \hline

            \textit{Grammar \newline Similarity} &
            Calculates the similarity between the schemas of the G.T and Generated Viz. &
            \small{
            Jaccard Similarity between the sets of keys in schema of the two visualizations
            }
             &
            \faCogs \\ 

            \hline

            \textit{Data \newline Mapping} &
            Assesses the data correctness of the Generated Viz. against G.T.
            &
           
            \begin{equation}
                \left\{\begin{matrix}
                100  &  if Data_{\text{G.T}} = Data_{\text{Gen}} \\ 
                \frac{\text{Matched Data Count}}{\text{Total Data Keys}} \times 100  & ow.
                \end{matrix}\right.
                \notag
            \end{equation}
             &
            \faCogs \\

            \hline

            \textit{Mark \newline Correctness} &
            Outputs 1 if mark type of Generated Viz. is same as that of G.T, 0 if not. &
            \small{
                \begin{equation}
                    \text{if Mark}_{\text{Gen}} = \text{Mark}_{\text{G.T}} \ \text{return} \ 1 \ \text{else} \ 0
                    \notag
                \end{equation}
            }
             &
            \faCogs \\

            \hline

            \textit{Axes \newline Quality} &
            Measures the quality of axes of visualization based on the axes in G.T and context of user query.
            &
            Strict comparison of encoding properties (axes) and human interpretation
            &
            \faUserCog \\

            \hline

            \textit{Color \newline Mapping} &
            Quality of the color mapping is measured against type of data in G.T and user query &
            Human Interpretation and Evaluation &
            \faUser \\

            \hline

            \textit{Image \newline Similarity} &
            Returns pixel level similarity/distance between the two visualizations &
            \small{
                \begin{equation}
                    \text{Image Similarity}_{\text{LPIPS}} = 100 - \text{LPIPS Distance(G.T, Gen)}
                    \notag
                \end{equation}  
                \begin{equation}
                    \text{Image Similarity}_{\text{SSI}} = \text{SSI(G.T, Gen)} \times 100
                    \notag
                \end{equation}
            }
            &
            \faCogs \\

            \hline

            \textit{Perceptual \newline Similarity} &
            Perceptual accuracy is determined in this level &
            Human Interpretation and Evaluation &
            \faUser \\

            \hline

            \textit{Visualization \newline Literacy} &
            Assesses the visualization on the basis of best standards of visualizations &
            Human Interpretation and Evaluation &
            \faUser \\

            \hline

            \textit{Significance} &
            Measures the insightfulness of the visualization &
            Human Interpretation and Evaluation &
            \faUser \\

            \hline

        \end{tabular}
    \caption{Enhancing EvaLLM Stack Levels: Approach and Design choice for evaluation at each stack level. In human terms, \faCogs symbolizes a fully automatic approach, \faUserCog refers to the hybrid approach, and \faUser represents a fully human evaluation approach.   }
    \label{tab:evallm_levels_table}
\end{table*}


To prove the EvaLLM stack's implementation and support the automatic and manual evaluation activities that the stack implies, we developed an online evaluation platform based on it.
The platform's design supports an incremental workflow where a generated visualization passes through it, and it is evaluated on the different EvaLLM levels using a selection of possible evaluation metrics implemented at each level. To implement the platform, we report design choices at each level of the EvaLLM stack in Table~\ref{tab:evallm_levels_table}. Moreover, the assessments are categorized into two distinct methods: automated processes, where the platform furnishes direct scoring, and manual activities, wherein the platform facilitates the collaboration of multiple users (referred to as assessors) engaged in manual evaluation tasks.
The evaluation platform uses a Python code base and an Angular frontend. The instance we provide is available at the following link: \url{https://github.com/lucapodo/evallm.git}. It represents an exemplification of how EvaLLM can be put into practice, 
It has been effectively used to populate the evaluation data discussed for the presented use cases in Section~\ref{sec:usecases}.

\begin{figure*}[t!]
    \centering
    \includegraphics[width=\linewidth]{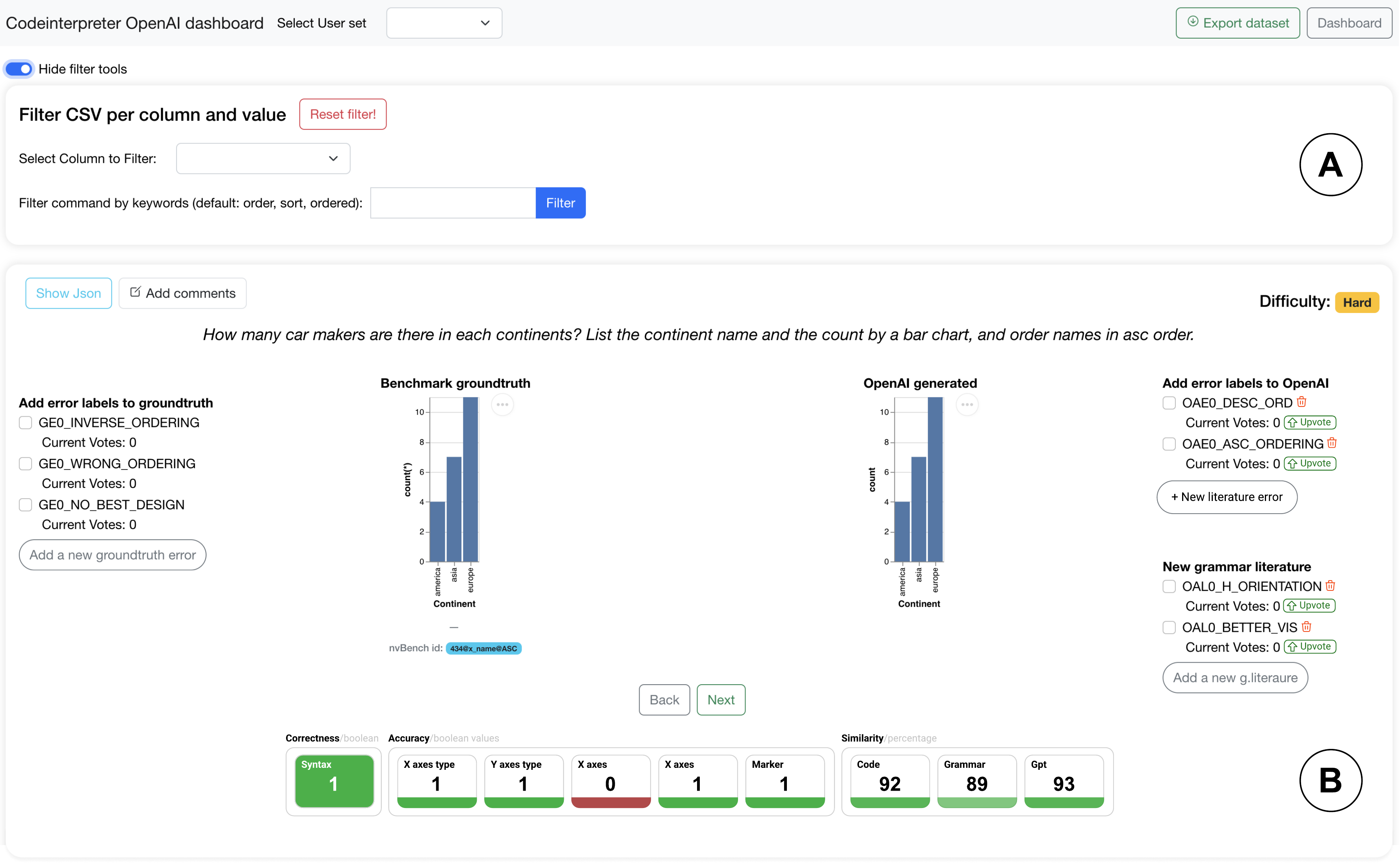}
    \caption{EvaLLMStack: concept evaluation stack for evaluating the LLM-generated visualizations}
    \label{fig:dashboard}
\end{figure*}


\subsection{Platform Interface}
\label{sec:dashboard}

An overview of the EvaLLM platform is visible in Figure~\ref{fig:dashboard}.
The visual design organizes the view into two main areas: the data and filter environment (Figure~\ref{fig:dashboard}.A) and the visual evaluation environment (Figure~\ref{fig:dashboard}.B).
Focusing on the former lets the human assessor review and explore the sets of experiments run through the LLM under evaluation. The assessor can upload the results of the tests run for the LLM into the system and study them. Through this view, specific filters can be used to retrieve particular instances, or a filter search bar allows retrieving them by keywords present in the tested prompts. This design choice has been taken to decouple the automatic testing phase from the human-based evaluation and analysis of results, as the first phase could be pretty long and could be run in a different environment (e.g., Google Colab, local machine, AWS Sagemaker).
The visual evaluation environment allows instead to explore single instances of the run experiments for manual inspection: the rationale is permitting a pool of assessors to (a) review the set of automatic scores computed at the different levels of the EvaLLM stack (e.g., evaluate the syntax correctness, or the mark correctness), and (b) allowing manual labeling for the stack level for which manual activities have been designed (e.g., evaluating the significance of a generated visualization, or the adherence to visualization literacy).
For this reason, it reports the rendered visualizations coming from the ground truth (on the left) and the one generated from the tested LLM (on the right) in the central area. Above them is reported the text of the original prompt used for generating the visualization to provide contextual information for the human assessor. Where present in the ground truth, the task's degree of difficulty is reported, too, along with additional contextual information.
The human assessor can manually review and inspect the differences between the two visualizations and, related to the mostly manual levels of the EvaLLM stack (e.g., the presentation layer), annotate potential anomalies/errors discovered using the platform. The inserted annotations report the error's name and the number of votes that different human assessors have expressed for that error. An assessor can even insert a new type of error or retain the ones already inserted by other assessors to minimize duplication and ambiguity.
The interface to execute this operation is at the far side of the generated visualization and the ground truth. The explanation for supporting this task and the ground truth could seem counter-intuitive. Still, we chose to keep it as, during our preliminary test with NVbench data, we were confronted multiple times with situations in which even the ground truth seemed to present some minor errors on perceptual quality or visualization literacy levels. In contrast, the LLM-based one did not give these errors. To avoid penalizing this case for not adhering entirely to the ground truth, we choose to allow manual labeling for them.
Finally, in the bottom part of the interface, the scores for the automatic evaluation functions defined for EvaLLM levels can be reviewed. They are organized using a small multiples visualization technique, where each score is reported, and the name, numerical value, and simple progress bar are colored according to how good the score is.
A pool of human assessors can independently traverse the tested cases, review the automatic scores, complete the manual labeling, be supported in commuting final consensus, and complete the evaluation campaign for one LLM or a set of LLMs.
When the activity is finished, all the data can be exported as a JSON file for further analysis or communication.

This paper investigated the elements to consider for evaluating in a comprehensive and fine-grained way an LLM-based generated visualization. Those elements were condensed and structured formally into the proposed EvaLLM stack, the first proposal targeted at LLMs.
To make the stack applicable and useful for instructing benchmarking activities, we contributed a web-based platform that can support implementing, testing and reviewing the results of the evaluation using EvaLLM. Additionally, it also supports the implementation of manual labelling from multiple assessors.
Two use cases, based on Code Interpreter and Llama-70b show the benefits and results obtainable by using the proposed evaluation stack and platform.

In future works, we plan to extend the experimental setup to more samples from the used datasets, additional datasets and models. We also foresee providing a survey on existing techniques that in visualization and visual analytics literature can support the implementation of the evaluation at the different levels of EvaLLM.

\section{Use cases}
\label{sec:usecases}

This section shows the advantages that using EvaLLM enables through two practical use cases. Our chosen scenarios involve evaluating the capabilities of GPT-3.5-Turbo Codeinterpreter and Llama2-70b \footnote{\url{https://platform.openai.com/docs/models}, \url{https://ai.meta.com/llama/}} to generate VegaLite visualizations from a given dataset, and a user query. The dataset is a reduced version of the well-known NVBench dataset that remains representative of its visualization general characteristics. 


\subsection{Methodology}
\label{sec:meth}
The primary objective of these use cases is to provide an example of how to assess the capabilities of Llama2-70b and GPT-3.5-turbo, two real and used LLMs, in generating a visualization grammar based on the input of a dataset and a user query. The interaction with the LLMs is facilitated through the zero-shot prompt engineering technique in both scenarios. Zero-shot prompting~\cite{xian2018zero} entails querying the LLM to produce task-specific output without providing any examples or additional context in input. An example is to ask the model to generate a VegaLite visualization based solely on a dataset and a user utterance.\\ 
This approach takes advantage of the inherent knowledge embedded in the LLM during the training phase to solve the proposed task. 
More sophisticated approaches, but requiring more effort, may involve few-shot learning or fine-tuning the model on the specific task.
To generate responses, the models are queried with a user utterance and a dataset mapped into a predefined prompt template. In the first use case, a code wrapper is employed to facilitate the model's processing of the visualization as code, subsequently returning it as a grammar structure. Conversely, the second use case involves the raw model without any intermediary wrapper.
Irrespective of the interaction type, the models share a common objective: returning the most pertinent VegaLite grammar visualization for the provided user utterance and dataset. The quality of the generated visualization is evaluated against ground truth using the EvaLLM platform to produce the final evaluation outcome.
The methodology employs an automatic analysis for the code and representation layer, complemented by a human-based evaluation focusing on the presentation, and application layers. Following classic benchmarks for LLM performances, qualitative aggregate results are summarized using a radar chart, considering only the automatic measure score

\subsection{Setup}
As outlined in the methodology, the setup for both use cases is generally consistent, differing primarily in the model employed and the utilization of a code model wrapper.
The experimental models include Meta's Llama2-70b and OpenAI's GPT-3.5-turbo. Specifically, for Llama2-70b, we utilized the plain model without any wrapper. In contrast, for GPT-3.5-turbo, we employed the open-source CodeInterpreterAPI~\cite{CodeInterpreterAPI}. This wrapper seamlessly integrates with the OpenAI GPT API, offering a fully integrated interface. 
For Llama, we opted for the cloud service DeepInfra ~\cite{deepinfra}, allowing us to query the model through API calls and maintain the model's operation on A100 GPUs.
To direct the models towards the defined objectives, we used the nvBench dataset ~\cite{luo2021nvbench}, considering a qualitative subset of 50 instances. 
As discussed, nvBench serves as a benchmark dataset for Natural Language to Visualization (NL2VIS) applications, providing instances that include user utterances for visualization generation, related datasets, and the corresponding visualizations in VegaLite.
Each instance passed to the models is reshaped using the prompt template based on the one proposed by Alpaca~\cite{taori2023alpaca}.
Subsequently, we evaluated the code layer and representation quality measures using an automatic approach, ensuring the essential structure of Json Vega-Lite matched. Human reviewers were also engaged for the presentation and application layers. 
More in detail, each level of EvaLLM was managed using the following approach:
\begin{itemize}
    \item Syntactic correctness level: Altair was employed to verify the correctness of the JSON structure.
    \item Code similarity: The \cite{pycodesimilar} Python library was utilized.
    \item Grammar similarity: A custom function was defined to compare two JSON structures without involving the values for the keys.
    \item Data mapping, mark correctness, axes quality: Custom JSON functions were implemented to compare corresponding fields between the ground truth and predictions.
    \item Color, image, perceptual, visualization, and significance levels: These aspects were evaluated based on a human reviewer approach, involving two users who reviewed the generated visualizations and provided labels corresponding to potential errors generated by the models.
\end{itemize}

The full setup is summarized in Table~\ref{tab:evallm_levels_table}.
All the code and the platform will be accessible on this link \url{https://github.com/lucapodo/evallm},for running the experiment.

\subsection{Use case 1: Evaluating Codeinterpreter}
\begin{figure}[!h]
    \centering
    \includegraphics[width=0.85\linewidth]{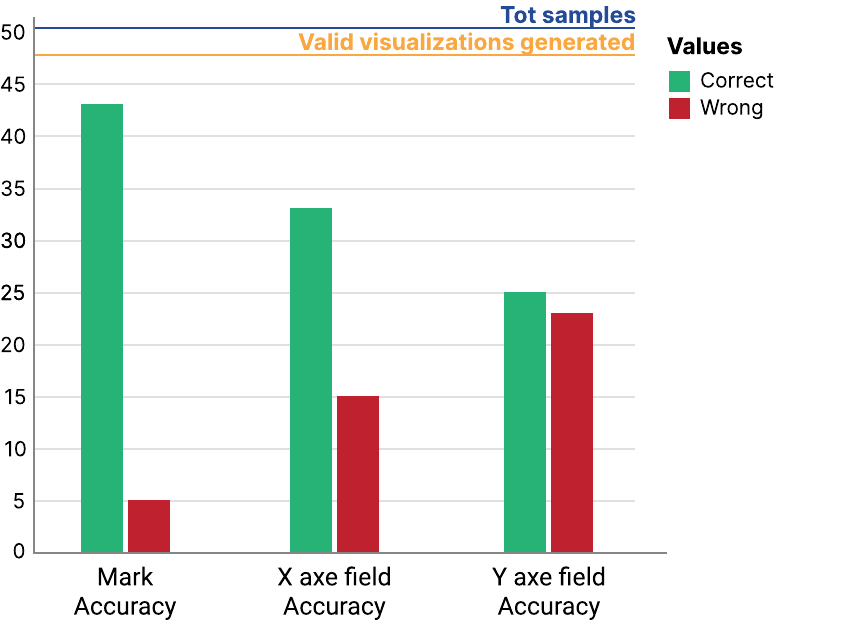}
    \caption{GPT-3.5-Turbo performance on 50 nvBench samples along mark and axes fields accuracy.}
    \label{fig:gpt results}
\end{figure}

In our investigation, we initially delved into the nvBench dataset, focusing our evaluation on a representative subset of 50 instances randomly sampled from the full dataset. 
Focusing on code and representation layers, the analysis of the results indicates that the model successfully generated 48 valid visualizations out of the 50 proposed samples from the dataset. These visualizations are deemed valid due to their compatibility with the VegaLite rendering. Figure~\ref{fig:gpt results} provides an overview of the model's performance for three dimensions: mark type, x-axis field, and y-axis field accuracy. It can be observed how the model exhibited an adequate mark accuracy rate, with only 5 instances out of the 48 valid visualizations incurring erroneous generation.
However, nuanced scrutiny of the results uncovered a relative weakness in the model's capability to discern the appropriate columns from the dataset when populating the x-axis and y-axis fields. Notably, the model's proficiency waned, especially concerning the accurate selection of the y-axis field, where the model correctly selected 25 out of 48 instances, compared to its performance with the x-axis where the model improved the performance by selecting 33 out of 48 instances, exhibiting a lower accuracy for the former.


\begin{figure*}[!h]
    \centering
    \includegraphics[width=\textwidth]{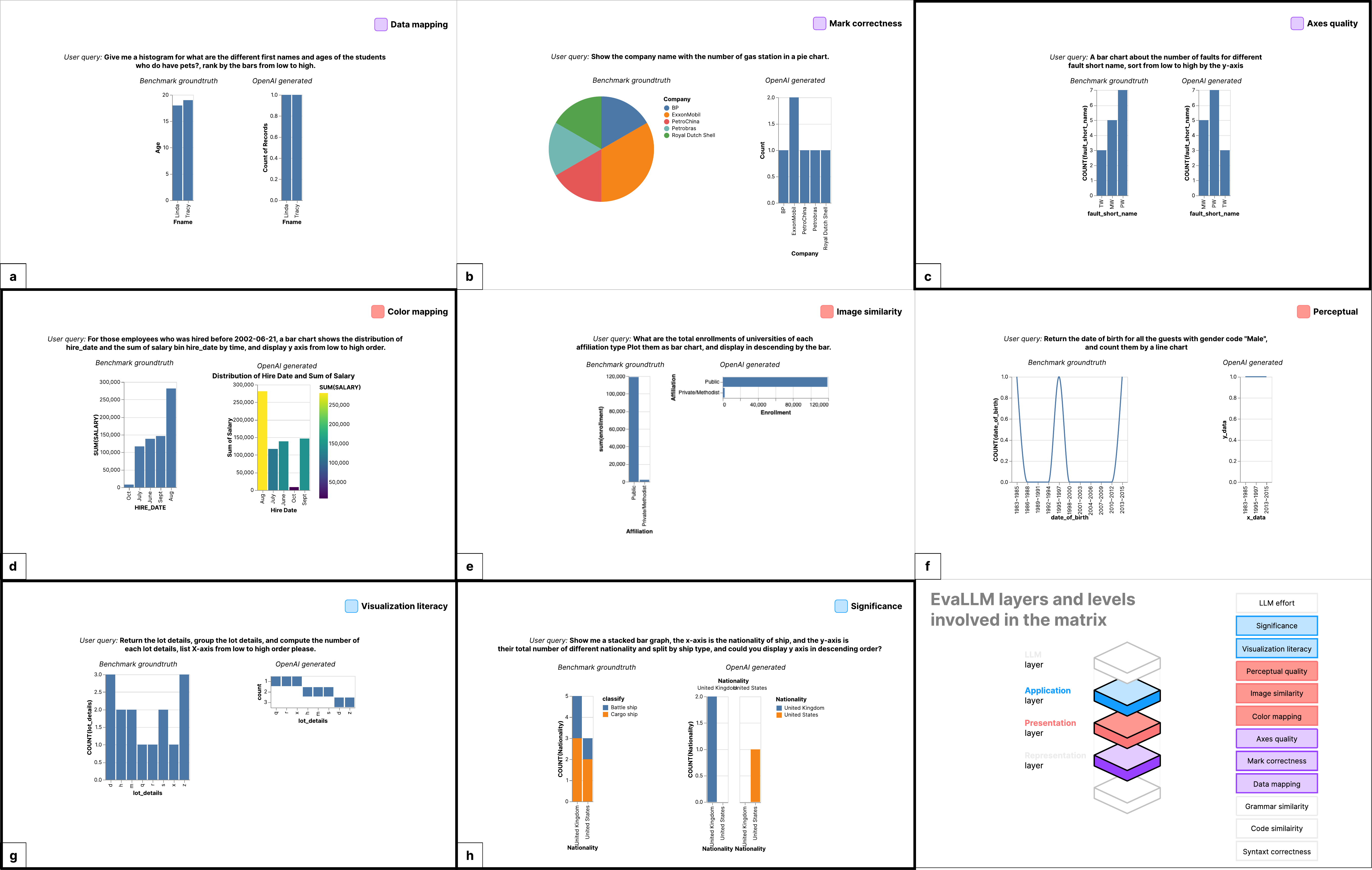}
    \caption{Examples of wrong generation by GPT-3.5 split by EvaLLM levels. Where the levels are purely numerical, the example is not reported.}
    \label{fig:gpt_grid}
\end{figure*}

Looking now at presentation and application layers, the samples have been analyzed using the platform, involving a human evaluator to label errors. 
This analysis has returned the identification of four classes of error that were not detected by the automated approach: 
\begin{itemize}
        \item \textbf{Missed Ordering Error:} The model misinterprets the user's explicit ordering instruction in the query. For instance, as illustrated in Figure~\ref{fig:gpt_grid}-c, despite a clear user directive (\textit{A bar chart about the number of faults for different fault short names, sort from low to high by the y-axis}), the model fails to adhere to the specified ordering. The second type is the absence of ordering specification in the query, which results in the model's inability to arrange the data in accordance with visualization literacy when it chooses a default sorting strategy.
        \item \textbf{Wrong Stacked Bar Chart:} Instances were observed where the model encountered challenges in accurately generating the VegaLite specification for a stacked bar chart, as exemplified in Figure ~\ref{fig:gpt_grid}-h. Instead of stacking the data as expected in the ground truth, the model diverged and fragmented the chart into multiple sub-charts. Moreover, each sub-chart presents only one value, missing the complementary one.
        \item \textbf{Visualization Hallucination:} This class of error manifests when the model invents an entirely different visual design for a well-known visualization technique requested in the query. Figure ~\ref{fig:gpt_grid}-g and ~\ref{fig:gpt_grid}-h demonstrate a scenario where the model completely misinterprets a stacked bar chart, deviating from established norms, and creates new representations that do not seem to follow any established rationale (e.g., the two distant bars reported in ~\ref{fig:gpt_grid}-h).
        \item \textbf{Unnecessary Color coding:} Errors in this category involve the addition of color coding, gradients, or other color properties that contravene best practices in visualization. Figure~\ref{fig:gpt_grid}-d illustrates instances where the model introduces unnecessary colors, detracting from the visual clarity recommended by best practices.
\end{itemize}

\subsection{Use case 2: Evaluating Llama-70b}
\label{sec:llama70}



\begin{figure}[!h]
    \centering
    \includegraphics[width=0.85\linewidth]{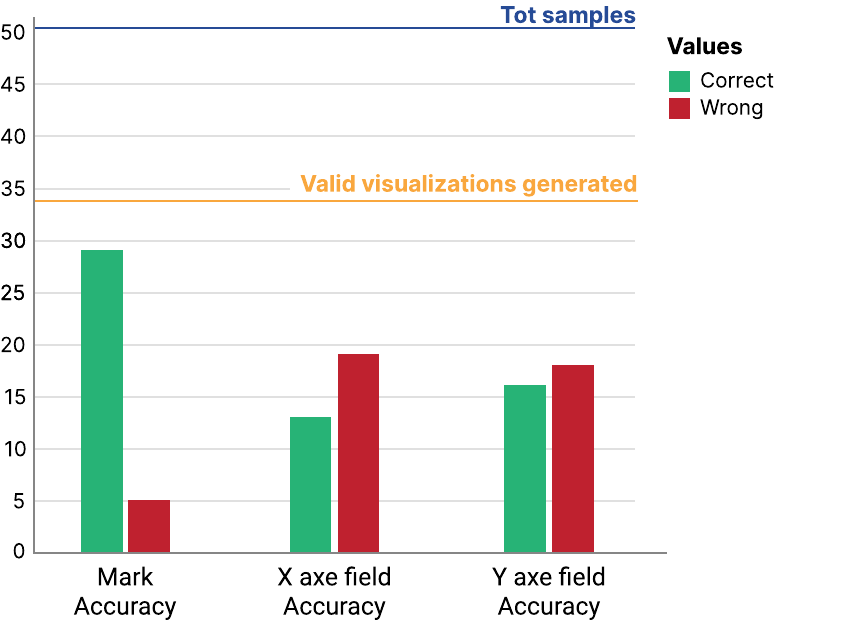}
    \caption{Llama2-70b performance on 50 nvBench samples along mark and axes fields accuracy.}
    \label{fig:llama results}
\end{figure}

\begin{figure*}[!h]
    \centering
    \includegraphics[width=\textwidth]{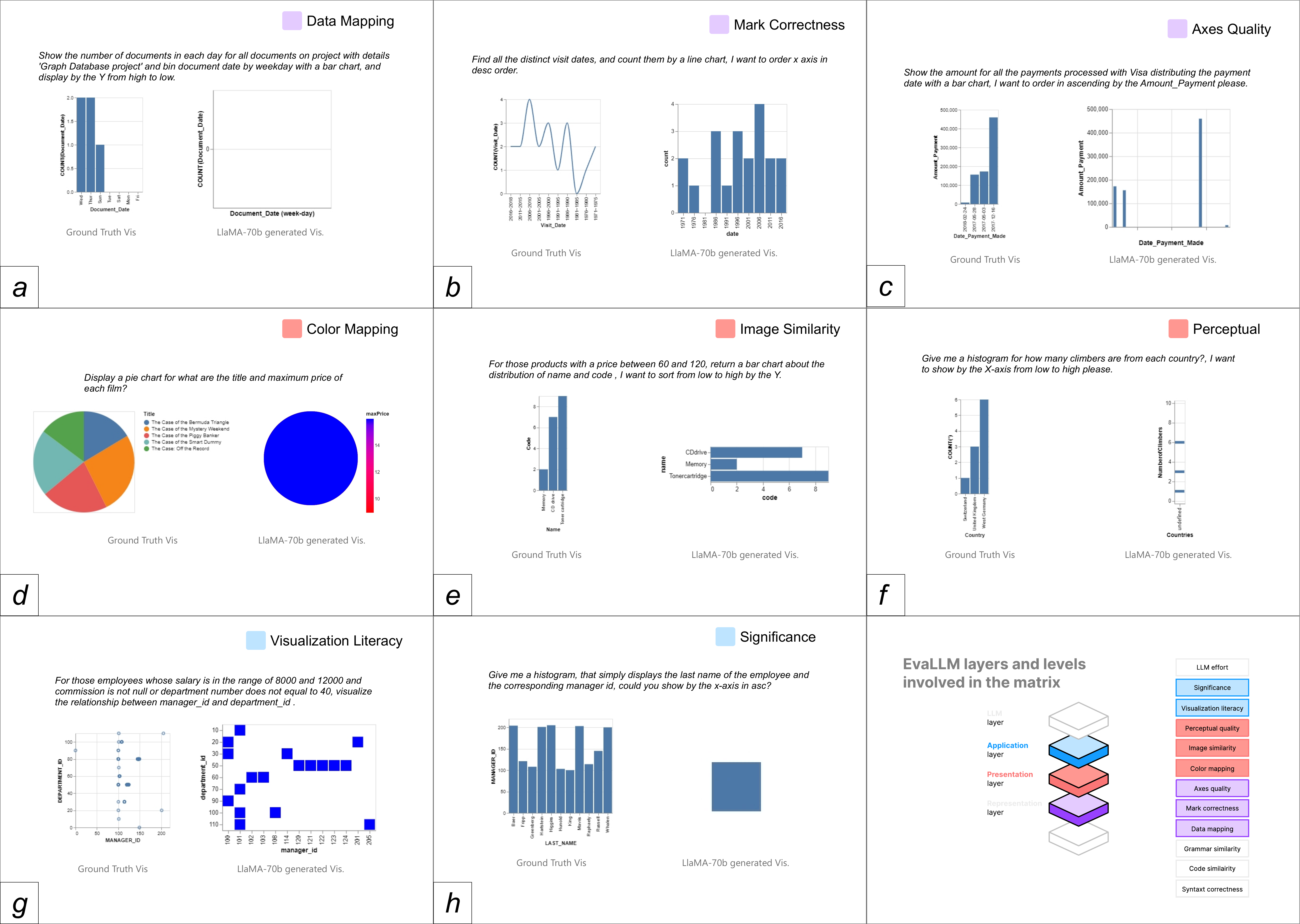}
    \caption{Examples of wrong generation by LlaMA2-70b split by EvaLLM levels. Where the levels are purely numerical, the example is not reported.}
    \label{fig:llama_grid}
\end{figure*}

This section illustrates the exploration of Meta LlaMA2-70b-chat ~\cite{touvron2023llama} as the second use case, with a focus on comparing it to GPT-3.5 Turbo~\cite{openai2023gpt} using again a subset of 50 instances from the NvBench dataset~\cite{luo2021nvbench}. It revealed interesting insights as depicted in Figure\ref{fig:llama results}: out of the 50 samples, 34 visualizations were successfully generated without any errors in adherence to the VegaLite schema. This is a worse result than what we experienced with ChatGPT 3.6-turbo CodeInterpreter, showing a worse capability of a non-fine-tuned model to support even the base level of code generation.
The performance of the LLM was evaluated across three dimensions for the representation layer: mark type accuracy, x-axis field accuracy, and y-axis field accuracy as demonstrated in Figure~\ref{fig:llama results}. Notably, there was a slight dip in mark type accuracy, with the LLM correctly identifying the mark type in 29 out of 34 visualizations, as opposed to 43 out of 48 by GPT-3.5 Turbo. This discrepancy suggests a similar difference in the models' capabilities even accounting for the lower number of tests.
Further analysis revealed a challenge for the LLM in handling largely structured prompts. Some visualizations generated were either blank or presented visual hallucinations, lacking any meaningful insight (see for reference the big blue square in Figure~\ref{fig:llama_grid} [h]). This limitation points to the model's struggle with certain types of input structures, affecting its ability to produce informative visualizations.
Moreover, it was observed that the LLM faced difficulties in correlating NL-instructions, data, and sorting techniques. This led to the generation of visualizations that were incorrectly sorted and disoriented, similar to what was observed for use case 1, highlighting a potential area for improvement in the model's comprehension and synthesis of complex instructions.
\begin{itemize}
    \item \textbf{Inability of Incorporation of Data Values}: While the automated performance evaluation revealed a major dip in the performance of the model, a following human evaluation uncovered a few more inabilities of the model. Such is the inability of the model to correctly understand and incorporate the data values in the visualization based on user query. In some cases, the model was observed generating no data values at all as can be seen in Figure~\ref{fig:llama_grid} [a,g]. A few other generated visualizations were found to have data values ignored but had data linked to a separate data file, which was quite strange behavior.

    \item \textbf{Largely Structured Prompts Ignored}: Investigating why the model was generating visualizations with no data values plotted at all unearthed some findings on the cause for the missing generation: in particular, in a few of the cases where the prompt structure was rather simple but with a lot of data, the model sent back as response the prompt and sometimes just blank strings without any text or warning of some kind.

    \item \textbf{Low Visualization Significance}: 
    In several visualizations generated by the model, there was little to no significance to the visualization in terms of the user query and data. Such examples can be seen, to different degrees, in Figure \ref{fig:llama_grid} [a,c,f,g,h]. Specifically, the visualization in Figure \ref{fig:llama_grid}-h is the most common visualization in terms of the lack of significance the model was found to be generating. In this aspect, LlaMA2-70b seems to perform significantly worse than GPT 3.5-turbo.

    \item \textbf{Incorrect or missing Sorting}: 
    In cases where the user query explicitly identified the nature of the sorting, the model was unable to understand it, resulting in unordered or incorrectly ordered data values. One instance of this can be seen in Figure ~\ref{fig:llama_grid}-e.
\end{itemize}

In conclusion, while the Llama2-70b demonstrated competence in generating valid visualizations, it exhibited challenges in code (generation), representation {(mark type accuracy), presentation (correlation between instructions and data sorting), and application(significance for structured prompts) layers. These findings provide valuable insights for future enhancements and optimizations in the development of LLM for visualization generation tasks.\\
\noindent We conclude the analysis of both use cases hinting at a potential use of EvaLLM in the form of a benchmark for analyzing and comparing the performances of multiple LLMs. Figure~\ref{fig:comparison} shows the result for the two tested models.
The evaluation is centered on five key dimensions: Mark Correctness, Data Mapping Correctness, Syntax Correctness, Grammar Similarity, and Code Similarity, covering in this example five levels and two layers of the stack. We choose to report just the automatic analyses given the explorative nature of the conducted use cases, which are not complete. 
This focused approach ensured a rigorous and measurable analysis, allowing for a concrete comparison of the LLMs' performance.


\begin{figure}[!h]
  \begin{subfigure}[b]{0.45\columnwidth}
    \includegraphics[width=\linewidth]{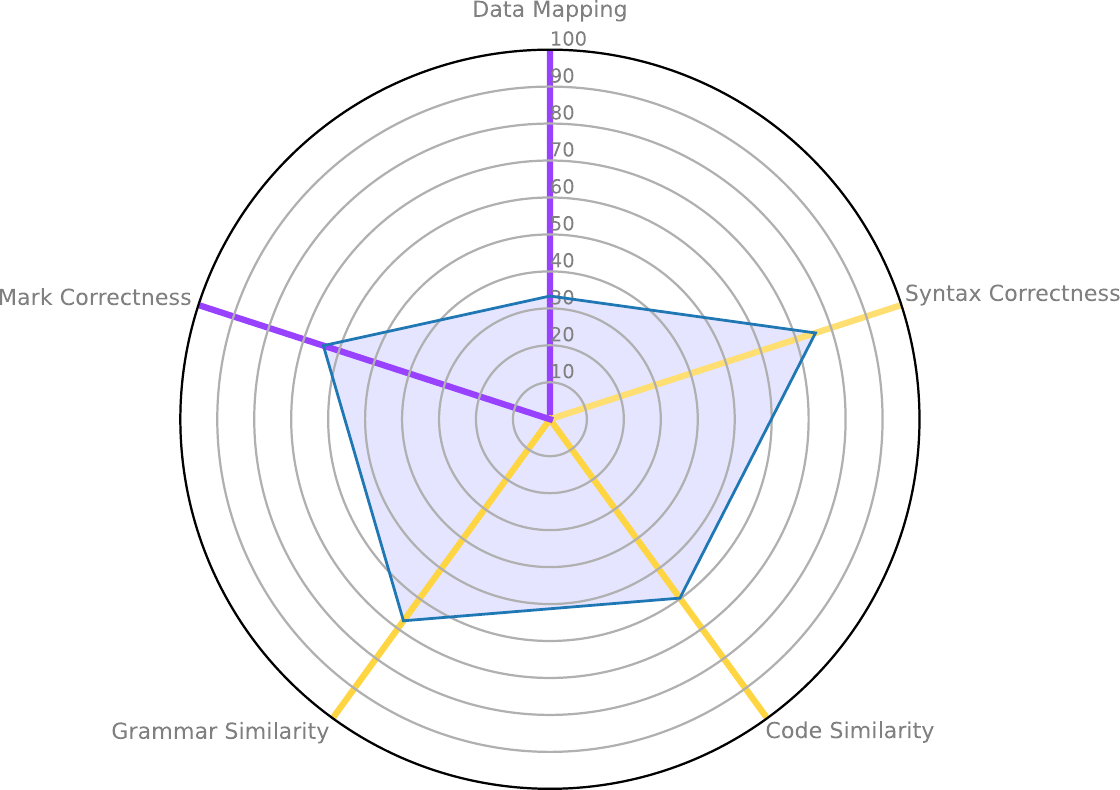}
    \caption{Llama2-70b}
    \label{fig:1}
  \end{subfigure}
  \hfill 
  \begin{subfigure}[b]{0.45\columnwidth}
    \includegraphics[width=\linewidth]{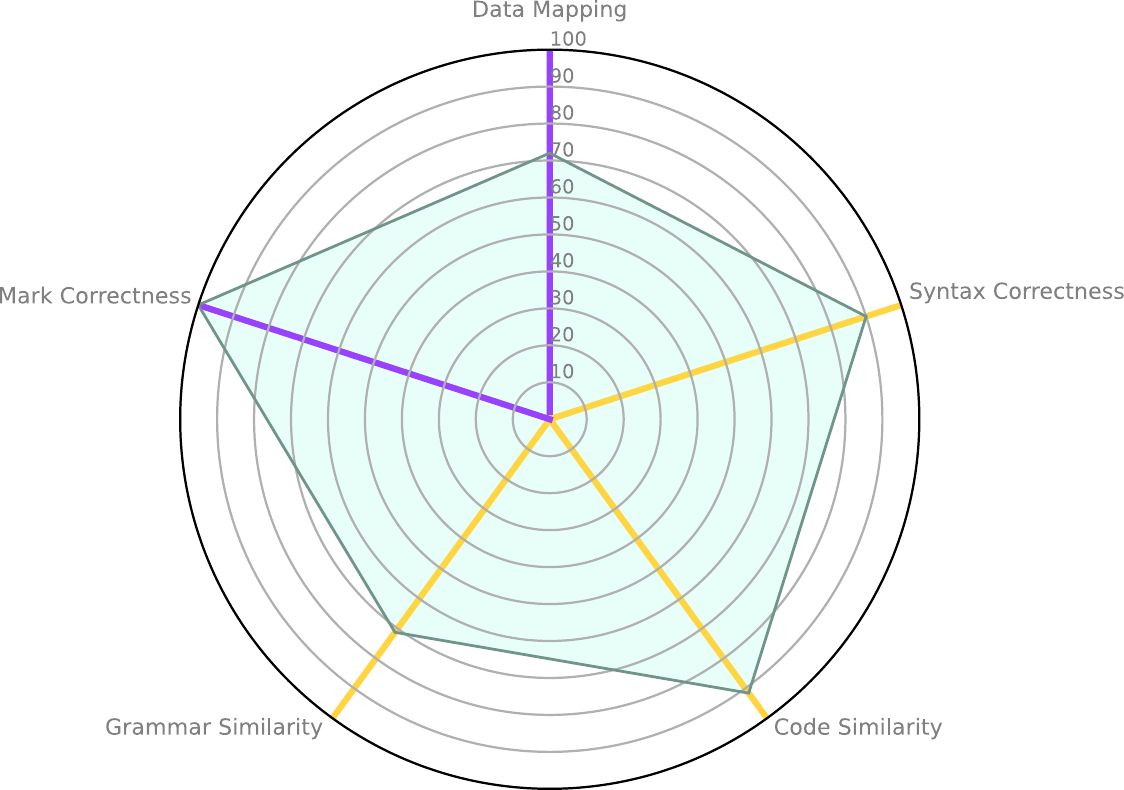}
    \caption{GPT-3.5 Turbo}
    \label{fig:2}
  \end{subfigure}
  \caption{Comparative analysis of the two tested LLM}
  \label{fig:comparison}
\end{figure}

\section{Threats to validity and Opportunities}
The primary objective of this work was to identify a comprehensive evaluation stack for a standard and multi-faceted assessment of LLMs for their use in visualization generation tasks. 
This objective focused the work on the generality and applicability of the proposed framework, more than on the evaluation of single LLMs.
The presented framework represents a first contribution to initiate a set of evaluation activities in the data visualization domain for this new technology and help researchers in quantitative comparisons of their proposed approaches.
The paper presents altogether a set of threats to its validity that can ask for further investigation, and that we report in the following.

\noindent\textbf{Limited number of experiments} This paper explores a limited number of samples from nvBench to demonstrate the EvaLLM applicability and benefits. While the chosen examples come from a state-of-the-art benchmarking dataset, their number is still on the low end. The reason behind this limitation is the low availability of free resources for testing many LLM models, resulting in the need for high-end computing machines or significant expenses even for a limited number of tests. We plan to integrate tests as future activity, exploiting the live nature of the contributed evaluation platform, and we foresee this limitation to be overcome as more open source models become available.


\noindent\textbf{Coverage of visualization techniques}
Among the used and existing benchmarking datasets, one limitation is represented by the narrow coverage of visualization techniques, with a high presence of simple visualizations (e.g., bar charts, pie charts, scatterplots) and a little to non-existant presence of more complex ones (e.g., RadViz, Sankey Diagrams, graphs, parallel coordinates, etc.). 
An essential enhancement would involve the creation of a more structured dataset incorporating complex visualizations such as geo maps or matrix-based visualizations, challenging LLMs to generate more sophisticated outputs. This expansion would better reflect the diverse landscape of visualization and foster a more comprehensive understanding of LLM capabilities and limitations. Expansion toward full dashboard evaluation could be followed too, according to what was recently proposed by Srinivasan and Setlur~\cite{Srinivasan2023} for natural language rule-based approaches.

On the other hand, EvaLLM can foster further research to overcome these limitations, and going beyond them in the following directions:

\noindent\textbf{Characterization of LLMs errors} EvaLLM stack can be leveraged to develop an extensive taxonomy of errors during the generation, relating them to the proposed layers and levels, encompassing a broader description of common mistakes in tested LLMs. This taxonomy can be utilized to assess existing and forthcoming models, gauge their evolution in visualization generation tasks, and inform the creation of an overview at the error level useful to hypothesize potential causes and mitigations. Even in our little exploration, we were already able to identify common types of errors shared by the two tested models.

\textbf{Inform comparative benchmarking on Visualization Capabilities of LLM}
EvaLLM establishes the groundwork for a standardized evaluation methodology and benchmarking tailored for the visualization task across different LLMs. Currently, existing quantitative methods lack the ability to effectively compare the performance of diverse models in generating visualizations. EvaLLM seeks to address this shortage by introducing an approach that can be seamlessly incorporated into quantitative comparative analyses, mirroring methodologies successfully employed for other generic tasks like math proficiency or conversational capabilities.

\section{Conlusion}
\label{sec:conclusion}
This paper investigated the elements to consider for evaluating in a comprehensive and fine-grained way an LLM-based generated visualization. Those elements were condensed and structured formally into the proposed EvaLLM stack, the first proposal targeted at LLMs.
To make the stack applicable and useful for instructing benchmarking activities, we contributed a web-based platform that can support implementing, testing and reviewing the results of the evaluation using EvaLLM. Additionally, it also supports the implementation of manual labelling from multiple assessors.
Two use cases, based on ChatGPT 3.5-turbo CodeInterpreter and Llama2-70b show the benefits and results obtainable by using the proposed evaluation stack and platform.

In future works, we plan to extend the experimental setup to more samples from the used dataset and consider additional datasets and models. We also foresee providing a survey on existing techniques that in visualization and visual analytics literature can support the implementation of the evaluation at the different levels of EvaLLM.

\bibliographystyle{eg-alpha-doi} 
\bibliography{bibliography}       

\newcommand{\etalchar}[1]{$^{#1}$}
\begin{thebibliography}{\uppercase{FADB{\etalchar{*}}22}}

\bibitem[AC17]{appalaraju2017image}
\textsc{Appalaraju S., Chaoji V.}:
\newblock Image similarity using deep cnn and curriculum learning.
\newblock \emph{arXiv preprint arXiv:1709.08761} (2017).

\bibitem[alt]{altairAltairDiscover}
{A}ltair | {D}iscover {C}ontinuously. {A}dvance {I}nfinitely - {O}nly
  {F}orward. --- altair.com.
\newblock \url{https://altair.com/}.
\newblock [Accessed 18-12-2023].

\bibitem[Bcnt]{D3js}
\textsc{Bostock M., contributors}:
\newblock D3.js - data-driven documents, 2009--present.
\newblock Accessed: \today.
\newblock URL: \url{https://d3js.org/}.

\bibitem[BO23]{Battle2023}
\textsc{Battle L., Ottley A.}:
\newblock What do we mean when we say “insight”? a formal synthesis of
  existing theory.
\newblock \emph{IEEE Transactions on Visualization and Computer Graphics}
  (2023), 1--14.
\newblock \href {https://doi.org/10.1109/TVCG.2023.3326698}
  {\path{doi:10.1109/TVCG.2023.3326698}}.

\bibitem[BRBF14]{boy2014principled}
\textsc{Boy J., Rensink R.~A., Bertini E., Fekete J.-D.}:
\newblock A principled way of assessing visualization literacy.
\newblock \emph{IEEE transactions on visualization and computer graphics 20},
  12 (2014), 1963--1972.

\bibitem[CCE{\etalchar{*}}18]{clark2018think}
\textsc{Clark P., Cowhey I., Etzioni O., Khot T., Sabharwal A., Schoenick C.,
  Tafjord O.}:
\newblock Think you have solved question answering? try arc, the ai2 reasoning
  challenge.
\newblock \emph{arXiv preprint arXiv:1803.05457} (2018).

\bibitem[CHL{\etalchar{*}}22]{chung2022scaling}
\textsc{Chung H.~W., Hou L., Longpre S., Zoph B., Tay Y., Fedus W., Li Y., Wang
  X., Dehghani M., Brahma S., et~al.}:
\newblock Scaling instruction-finetuned language models.
\newblock \emph{arXiv preprint arXiv:2210.11416} (2022).

\bibitem[CL23]{chiang2023can}
\textsc{Chiang C.-H., Lee H.-y.}:
\newblock Can large language models be an alternative to human evaluations?
\newblock \emph{arXiv preprint arXiv:2305.01937} (2023).

\bibitem[CLM{\etalchar{*}}22]{chen2022nl2interface}
\textsc{Chen Y., Li R., Mac A., Xie T., Yu T., Wu E.}:
\newblock Nl2interface: Interactive visualization interface generation from
  natural language queries.
\newblock \emph{arXiv preprint arXiv:2209.08834} (2022).

\bibitem[CTJ{\etalchar{*}}21]{chen2021evaluating}
\textsc{Chen M., Tworek J., Jun H., Yuan Q., Pinto H. P. d.~O., Kaplan J.,
  Edwards H., Burda Y., Joseph N., Brockman G., et~al.}:
\newblock Evaluating large language models trained on code.
\newblock \emph{arXiv preprint arXiv:2107.03374} (2021).

\bibitem[CW22]{chen2022pi2}
\textsc{Chen Y., Wu E.}:
\newblock Pi2: End-to-end interactive visualization interface generation from
  queries.
\newblock In \emph{Proceedings of the 2022 International Conference on
  Management of Data} (2022), pp.~1711--1725.

\bibitem[CZW{\etalchar{*}}23]{chen2023beyond}
\textsc{Chen Z., Zhang C., Wang Q., Troidl J., Warchol S., Beyer J., Gehlenborg
  N., Pfister H.}:
\newblock Beyond generating code: Evaluating gpt on a data visualization
  course.
\newblock \emph{arXiv preprint arXiv:2306.02914} (2023).

\bibitem[Dod02]{doddington2002automatic}
\textsc{Doddington G.}:
\newblock Automatic evaluation of machine translation quality using n-gram
  co-occurrence statistics.
\newblock In \emph{Proceedings of the second international conference on Human
  Language Technology Research} (2002), pp.~138--145.

\bibitem[DZ83]{day1983osi}
\textsc{Day J.~D., Zimmermann H.}:
\newblock The osi reference model.
\newblock \emph{Proceedings of the IEEE 71}, 12 (1983), 1334--1340.

\bibitem[FADB{\etalchar{*}}22]{finnie2022robots}
\textsc{Finnie-Ansley J., Denny P., Becker B.~A., Luxton-Reilly A., Prather
  J.}:
\newblock The robots are coming: Exploring the implications of openai codex on
  introductory programming.
\newblock In \emph{Proceedings of the 24th Australasian Computing Education
  Conference} (2022), pp.~10--19.

\bibitem[FBRK23]{firdous2023openai}
\textsc{Firdous F., Bashir S., Rufai S.~Z., Kumar S.}:
\newblock Openai chatgpt as a logical interpreter of code.
\newblock In \emph{2023 2nd International Conference on Edge Computing and
  Applications (ICECAA)} (2023), IEEE, pp.~1192--1197.

\bibitem[FNJL23]{fu2023gptscore}
\textsc{Fu J., Ng S.-K., Jiang Z., Liu P.}:
\newblock Gptscore: Evaluate as you desire.
\newblock \emph{arXiv preprint arXiv:2302.04166} (2023).

\bibitem[FXG{\etalchar{*}}20]{fu2020quda}
\textsc{Fu S., Xiong K., Ge X., Tang S., Chen W., Wu Y.}:
\newblock Quda: Natural language queries for visual data analytics.
\newblock \emph{arXiv preprint arXiv:2005.03257} (2020).

\bibitem[Fyr20]{pycodesimilar}
\textsc{Fyrestone}:
\newblock Pycode similar, 2020.
\newblock URL: \url{https://pypi.org/project/pycode-similar/}.

\bibitem[GDA{\etalchar{*}}15]{gao2015datatone}
\textsc{Gao T., Dontcheva M., Adar E., Liu Z., Karahalios K.~G.}:
\newblock Datatone: Managing ambiguity in natural language interfaces for data
  visualization.
\newblock In \emph{Proceedings of the 28th annual acm symposium on user
  interface software \& technology} (2015), pp.~489--500.

\bibitem[Gitnt]{GitHubCopilot}
\textsc{GitHub}:
\newblock Github copilot, 2021--present.
\newblock Accessed: \today.
\newblock URL: \url{https://github.com/features/copilot}.

\bibitem[HC21]{haq2021survey}
\textsc{Haq I.~U., Caballero J.}:
\newblock A survey of binary code similarity.
\newblock \emph{ACM Computing Surveys (CSUR) 54}, 3 (2021), 1--38.

\bibitem[HC23]{hong2023conversational}
\textsc{Hong M.-H., Crisan A.}:
\newblock Conversational ai threads for visualizing multidimensional datasets.
\newblock \emph{arXiv preprint arXiv:2311.05590} (2023).

\bibitem[HMB21]{hazoom2021text}
\textsc{Hazoom M., Malik V., Bogin B.}:
\newblock Text-to-sql in the wild: a naturally-occurring dataset based on stack
  exchange data.
\newblock \emph{arXiv preprint arXiv:2106.05006} (2021).

\bibitem[HPM{\etalchar{*}}23]{hull2023visgrader}
\textsc{Hull M., Pednekar V., Murray H., Roy N., Tung E., Routray S., Guerin
  C., Chen J., Wang Z.~J., Lee S., et~al.}:
\newblock Visgrader: Automatic grading of d3 visualizations.
\newblock \emph{IEEE Transactions on Visualization and Computer Graphics}
  (2023).

\bibitem[HQS{\etalchar{*}}23]{hadi2023large}
\textsc{Hadi M.~U., Qureshi R., Shah A., Irfan M., Zafar A., Shaikh M.~B.,
  Akhtar N., Wu J., Mirjalili S., et~al.}:
\newblock Large language models: a comprehensive survey of its applications,
  challenges, limitations, and future prospects.

\bibitem[Hun07]{Hunter:2007}
\textsc{Hunter J.~D.}:
\newblock Matplotlib: A 2d graphics environment.
\newblock \emph{Computing in Science \& Engineering 9}, 3 (2007), 90--95.
\newblock \href {https://doi.org/10.1109/MCSE.2007.55}
  {\path{doi:10.1109/MCSE.2007.55}}.

\bibitem[JSFL]{joshievaluating}
\textsc{Joshi A., Srinivas C., Firat E.~E., Laramee R.~S.}:
\newblock Evaluating the recommendations of llms to teach a visualiza-tion
  technique using bloom’s taxonomy.

\bibitem[KHL19]{kettunen2019lpips}
\textsc{Kettunen M., H{\"a}rk{\"o}nen E., Lehtinen J.}:
\newblock E-lpips: robust perceptual image similarity via random transformation
  ensembles.
\newblock \emph{arXiv preprint arXiv:1906.03973} (2019).

\bibitem[KLSC21]{kim2021data}
\textsc{Kim Y.-H., Lee B., Srinivasan A., Choe E.~K.}:
\newblock Data@ hand: Fostering visual exploration of personal data on
  smartphones leveraging speech and touch interaction.
\newblock In \emph{Proceedings of the 2021 CHI Conference on Human Factors in
  Computing Systems} (2021), pp.~1--17.

\bibitem[KMB23]{kim2023good}
\textsc{Kim N.~W., Myers G., Bach B.}:
\newblock How good is chatgpt in giving advice on your visualization design?
\newblock \emph{arXiv preprint arXiv:2310.09617} (2023).

\bibitem[KMK23]{katsogiannis2023survey}
\textsc{Katsogiannis-Meimarakis G., Koutrika G.}:
\newblock A survey on deep learning approaches for text-to-sql.
\newblock \emph{The VLDB Journal} (2023), 1--32.

\bibitem[LH18]{liu2018somewhere}
\textsc{Liu Y., Heer J.}:
\newblock Somewhere over the rainbow: An empirical assessment of quantitative
  colormaps.
\newblock In \emph{Proceedings of the 2018 CHI conference on human factors in
  computing systems} (2018), pp.~1--12.

\bibitem[LKK16]{lee2016vlat}
\textsc{Lee S., Kim S.-H., Kwon B.~C.}:
\newblock Vlat: Development of a visualization literacy assessment test.
\newblock \emph{IEEE transactions on visualization and computer graphics 23}, 1
  (2016), 551--560.

\bibitem[LPX{\etalchar{*}}18]{liu2018supporting}
\textsc{Liu W., Peng X., Xing Z., Li J., Xie B., Zhao W.}:
\newblock Supporting exploratory code search with differencing and
  visualization.
\newblock In \emph{2018 IEEE 25th International Conference on Software
  Analysis, Evolution and Reengineering (SANER)} (2018), IEEE, pp.~300--310.

\bibitem[Ltd23]{deepinfra}
\textsc{Ltd D.~P.}:
\newblock Deepinfra, 2023.
\newblock URL: \url{https://deepinfra.com/}.

\bibitem[LTL21a]{luo2021nvbench}
\textsc{Luo Y., Tang J., Li G.}:
\newblock nvbench: A large-scale synthesized dataset for cross-domain natural
  language to visualization task.
\newblock \emph{arXiv preprint arXiv:2112.12926} (2021).

\bibitem[LTL{\etalchar{*}}21b]{luo2021natural}
\textsc{Luo Y., Tang N., Li G., Tang J., Chai C., Qin X.}:
\newblock Natural language to visualization by neural machine translation.
\newblock \emph{IEEE Transactions on Visualization and Computer Graphics 28}, 1
  (2021), 217--226.

\bibitem[Mic20]{playwright}
\textsc{Microsoft}:
\newblock Playwright, 2020.
\newblock URL: \url{https://playwright.dev/}.

\bibitem[MS23]{maddigan2023chat2vis}
\textsc{Maddigan P., Susnjak T.}:
\newblock Chat2vis: Generating data visualisations via natural language using
  chatgpt, codex and gpt-3 large language models.
\newblock \emph{IEEE Access} (2023).

\bibitem[Mun09]{munzner2009nested}
\textsc{Munzner T.}:
\newblock A nested model for visualization design and validation.
\newblock \emph{IEEE transactions on visualization and computer graphics 15}, 6
  (2009), 921--928.

\bibitem[Nor06]{North2006}
\textsc{North C.}:
\newblock Toward measuring visualization insight.
\newblock \emph{IEEE Computer Graphics and Applications 26}, 3 (2006), 6--9.
\newblock \href {https://doi.org/10.1109/MCG.2006.70}
  {\path{doi:10.1109/MCG.2006.70}}.

\bibitem[NSD11]{North2011}
\textsc{North C., Saraiya P., Duca K.}:
\newblock A comparison of benchmark task and insight evaluation methods for
  information visualization.
\newblock \emph{Information Visualization 10}, 3 (2011), 162--181.
\newblock URL: \url{https://doi.org/10.1177/1473871611415989}, \href
  {http://arxiv.org/abs/https://doi.org/10.1177/1473871611415989}
  {\path{arXiv:https://doi.org/10.1177/1473871611415989}}, \href
  {https://doi.org/10.1177/1473871611415989}
  {\path{doi:10.1177/1473871611415989}}.

\bibitem[NSS20]{narechania2020nl4dv}
\textsc{Narechania A., Srinivasan A., Stasko J.}:
\newblock Nl4dv: A toolkit for generating analytic specifications for data
  visualization from natural language queries.
\newblock \emph{IEEE Transactions on Visualization and Computer Graphics 27}, 2
  (2020), 369--379.

\bibitem[Ope23]{openai2023gpt}
\textsc{OpenAI R.}:
\newblock Gpt-4 technical report.
\newblock \emph{arXiv} (2023), 2303--08774.

\bibitem[PPV23]{podo2023machine}
\textsc{Podo L., Prenkaj B., Velardi P.}:
\newblock Machine learning for visualization recommendation systems: Open
  challenges and future directions.
\newblock \emph{arXiv preprint arXiv:2302.00569} (2023).

\bibitem[RH08]{rouse2008understanding}
\textsc{Rouse D.~M., Hemami S.~S.}:
\newblock Understanding and simplifying the structural similarity metric.
\newblock In \emph{2008 15th IEEE international conference on image processing}
  (2008), IEEE, pp.~1188--1191.

\bibitem[SBT{\etalchar{*}}16]{setlur2016eviza}
\textsc{Setlur V., Battersby S.~E., Tory M., Gossweiler R., Chang A.~X.}:
\newblock Eviza: A natural language interface for visual analysis.
\newblock In \emph{Proceedings of the 29th annual symposium on user interface
  software and technology} (2016), pp.~365--377.

\bibitem[Scnt]{CodeInterpreterAPI}
\textsc{Shroominic, contributors}:
\newblock Codeinterpreter api.
\newblock GitHub Repository, 2023--present.
\newblock Accessed: \today.
\newblock URL: \url{https://github.com/shroominic/codeinterpreter-api}.

\bibitem[SG16]{szafir2016visualization}
\textsc{Szafir D.~A., Gleicher M.}:
\newblock Visualization-aware color design.
\newblock In \emph{EuroVis (Posters)} (2016), pp.~97--99.

\bibitem[SMWC23]{suh2023grammar}
\textsc{Suh A., Mosca A., Wu E., Chang R.}:
\newblock A grammar of hypotheses for visualization, data, and analysis, 2023.
\newblock \href {http://arxiv.org/abs/2204.14267} {\path{arXiv:2204.14267}}.

\bibitem[SNL{\etalchar{*}}21]{srinivasan2021collecting}
\textsc{Srinivasan A., Nyapathy N., Lee B., Drucker S.~M., Stasko J.}:
\newblock Collecting and characterizing natural language utterances for
  specifying data visualizations.
\newblock In \emph{Proceedings of the 2021 CHI Conference on Human Factors in
  Computing Systems} (2021), pp.~1--10.

\bibitem[Spe01]{spence2001information}
\textsc{Spence R.}:
\newblock \emph{Information visualization}, vol.~1.
\newblock Springer, 2001.

\bibitem[SS23a]{srinivasan2023bolt}
\textsc{Srinivasan A., Setlur V.}:
\newblock Bolt: A natural language interface for dashboard authoring.

\bibitem[SS23b]{Srinivasan2023}
\textsc{Srinivasan A., Setlur V.}:
\newblock {BOLT: A Natural Language Interface for Dashboard Authoring}.
\newblock In \emph{EuroVis 2023 - Short Papers} (2023), Hoellt T., Aigner W.,
  Wang B., (Eds.), The Eurographics Association.
\newblock \href {https://doi.org/10.2312/evs.20231035}
  {\path{doi:10.2312/evs.20231035}}.

\bibitem[SSZ{\etalchar{*}}23]{sun2023principle}
\textsc{Sun Z., Shen Y., Zhou Q., Zhang H., Chen Z., Cox D., Yang Y., Gan C.}:
\newblock Principle-driven self-alignment of language models from scratch with
  minimal human supervision.
\newblock \emph{arXiv preprint arXiv:2305.03047} (2023).

\bibitem[Sza17]{szafir2017modeling}
\textsc{Szafir D.~A.}:
\newblock Modeling color difference for visualization design.
\newblock \emph{IEEE transactions on visualization and computer graphics 24}, 1
  (2017), 392--401.

\bibitem[SZWJ22]{song2022rgvisnet}
\textsc{Song Y., Zhao X., Wong R. C.-W., Jiang D.}:
\newblock Rgvisnet: A hybrid retrieval-generation neural framework towards
  automatic data visualization generation.
\newblock In \emph{Proceedings of the 28th ACM SIGKDD Conference on Knowledge
  Discovery and Data Mining} (2022), pp.~1646--1655.

\bibitem[TCD{\etalchar{*}}23]{tian2023chartgpt}
\textsc{Tian Y., Cui W., Deng D., Yi X., Yang Y., Zhang H., Wu Y.}:
\newblock Chartgpt: Leveraging llms to generate charts from abstract natural
  language.
\newblock \emph{arXiv preprint arXiv:2311.01920} (2023).

\bibitem[TGZ{\etalchar{*}}23]{taori2023alpaca}
\textsc{Taori R., Gulrajani I., Zhang T., Dubois Y., Li X., Guestrin C., Liang
  P., Hashimoto T.~B.}:
\newblock Alpaca: A strong, replicable instruction-following model.
\newblock \emph{Stanford Center for Research on Foundation Models.
  https://crfm. stanford. edu/2023/03/13/alpaca. html 3}, 6 (2023), 7.

\bibitem[TMS{\etalchar{*}}23]{touvron2023llama}
\textsc{Touvron H., Martin L., Stone K., Albert P., Almahairi A., Babaei Y.,
  Bashlykov N., Batra S., Bhargava P., Bhosale S., et~al.}:
\newblock Llama 2: Open foundation and fine-tuned chat models.
\newblock \emph{arXiv preprint arXiv:2307.09288} (2023).

\bibitem[TX23]{tao2023mapping}
\textsc{Tao R., Xu J.}:
\newblock Mapping with chatgpt.
\newblock \emph{ISPRS International Journal of Geo-Information 12}, 7 (2023),
  284.

\bibitem[War19]{ware2019information}
\textsc{Ware C.}:
\newblock \emph{Information visualization: perception for design}.
\newblock Morgan Kaufmann, 2019.

\bibitem[WFH{\etalchar{*}}23]{white2023prompt}
\textsc{White J., Fu Q., Hays S., Sandborn M., Olea C., Gilbert H., Elnashar
  A., Spencer-Smith J., Schmidt D.~C.}:
\newblock A prompt pattern catalog to enhance prompt engineering with chatgpt.
\newblock \emph{arXiv preprint arXiv:2302.11382} (2023).

\bibitem[WIL{\etalchar{*}}23]{wu2023bloomberggpt}
\textsc{Wu S., Irsoy O., Lu S., Dabravolski V., Dredze M., Gehrmann S.,
  Kambadur P., Rosenberg D., Mann G.}:
\newblock Bloomberggpt: A large language model for finance.
\newblock \emph{arXiv preprint arXiv:2303.17564} (2023).

\bibitem[WSM{\etalchar{*}}18]{wang2018glue}
\textsc{Wang A., Singh A., Michael J., Hill F., Levy O., Bowman S.~R.}:
\newblock Glue: A multi-task benchmark and analysis platform for natural
  language understanding.
\newblock \emph{arXiv preprint arXiv:1804.07461} (2018).

\bibitem[WWS{\etalchar{*}}22]{wei2022chain}
\textsc{Wei J., Wang X., Schuurmans D., Bosma M., Xia F., Chi E., Le Q.~V.,
  Zhou D., et~al.}:
\newblock Chain-of-thought prompting elicits reasoning in large language
  models.
\newblock \emph{Advances in Neural Information Processing Systems 35} (2022),
  24824--24837.

\bibitem[WYKN20]{wang2020generalizing}
\textsc{Wang Y., Yao Q., Kwok J.~T., Ni L.~M.}:
\newblock Generalizing from a few examples: A survey on few-shot learning.
\newblock \emph{ACM computing surveys (csur) 53}, 3 (2020), 1--34.

\bibitem[XLS17]{xu2017sqlnet}
\textsc{Xu X., Liu C., Song D.}:
\newblock Sqlnet: Generating structured queries from natural language without
  reinforcement learning.
\newblock \emph{arXiv preprint arXiv:1711.04436} (2017).

\bibitem[XLSA18]{xian2018zero}
\textsc{Xian Y., Lampert C.~H., Schiele B., Akata Z.}:
\newblock Zero-shot learning—a comprehensive evaluation of the good, the bad
  and the ugly.
\newblock \emph{IEEE transactions on pattern analysis and machine intelligence
  41}, 9 (2018), 2251--2265.

\bibitem[ZCG{\etalchar{*}}23]{zhong2023agieval}
\textsc{Zhong W., Cui R., Guo Y., Liang Y., Lu S., Wang Y., Saied A., Chen W.,
  Duan N.}:
\newblock Agieval: A human-centric benchmark for evaluating foundation models.
\newblock \emph{arXiv preprint arXiv:2304.06364} (2023).

\bibitem[ZHB{\etalchar{*}}19]{zellers2019hellaswag}
\textsc{Zellers R., Holtzman A., Bisk Y., Farhadi A., Choi Y.}:
\newblock Hellaswag: Can a machine really finish your sentence?
\newblock \emph{arXiv preprint arXiv:1905.07830} (2019).

\end{thebibliography}


%
%

\end{document}